\documentclass[10pt, nolongbibliography, groupedaddress, showkeys, amsmath, amssymb, eqsecnum,aps,prd, nofootinbib]{revtex4-2}

\usepackage[dvipsnames]{xcolor}
\usepackage[colorlinks=true, allcolors=Blue]{hyperref}

\usepackage{tensor, mathtools, bm, dsfont} 
\usepackage{tabularray}
\UseTblrLibrary{booktabs}

\usepackage{graphicx} 

\begin{document}                                   
\title{Equivalence of first and second order formulations of the Einstein-Hilbert theory}
\begin{abstract}
We derive a general relation between the background effective actions, which directly proves that the two formulations of the Einstein-Hilbert theory with background fields are equivalent at the quantum level. This basic result has been substantiated  in a general background gauge, by explicit calculations at one-loop order of the corresponding counterterm Lagrangians. 
\end{abstract}

\author{F. T. Brandt}  
\email{fbrandt@usp.br}
\affiliation{Instituto de F\'{\i}sica, Universidade de S\~ao Paulo, S\~ao Paulo, SP 05508-090, Brazil}

\author{J. Frenkel}
\email{jfrenkel@if.usp.br}
\affiliation{Instituto de F\'{\i}sica, Universidade de S\~ao Paulo, S\~ao Paulo, SP 05508-090, Brazil}

 \author{S. Martins-Filho}   
\email{sergiomartinsfilho@usp.br}
\affiliation{Instituto de F\'{\i}sica, Universidade de S\~ao Paulo, S\~ao Paulo, SP 05508-090, Brazil}

\author{D. G. C. McKeon}
\email{dgmckeo2@uwo.ca}
\affiliation{
Department of Applied Mathematics, The University of Western Ontario, London, Ontario N6A 5B7, Canada}
\affiliation{Department of Mathematics and Computer Science, Algoma University,
Sault Ste.~Marie, Ontario P6A 2G4, Canada}

\date{\today}

\maketitle

\section{Introduction}
The first order formulation of gauge field theories involves Lagrangians that are linear in time derivatives, unlike the second order formulation, which features quadratic time derivative terms. In addition to the original fields, the first order approach introduces independent fields---such as the affine connection in gravity---which can lead to significant simplifications in the structure of the interaction vertices. For instance, the first order formulations of the Yang–Mills and Einstein–Hilbert (EH) theories contain only a simple cubic interaction, whereas the second order versions typically involve a large number of more complex terms. This structural   improvement     makes the first order approach more appropriate for quantum calculations \cite{buchbinder:1983a, McKeon:1994ds, Kalmykov:1994yj, Martellini:1997mu, Brandt:2015nxa, Brandt:2016eaj, McKeon:2020lqp, Brandt:2020vre}.

Such formulations   and their equivalence have been investigated from many points of view, both without \cite{Okubo:1979gt, Taubes:1979ps, Castellani:1981ue, Tseytlin:1981ks, Accardi:1997ps, Kiriushcheva:2006gp, Andrasi:2007dk, McKeon:2010nf, Dadhich:2012htv, lopez:2014, Frenkel:2017xvm, Benisty:2018fgu, Tenkanen:2020dge, McKeon:2020lqp, Brandt:2020vre, Lavrov:2021pqh, Gallagher:2022kvv, Rigouzzo:2023sbb} and with background field quantization \cite{buchbinder:1985, Alvarez:2017cvx, Brandt:2018wxe, Brandt:2024rsy, Brandt:2024kvs}. In the background field method \cite{Abbott:1980hw, weinberg:2000}, gauge  invariance is preserved at each stage by splitting the fields into classical background and quantum parts. It has been observed that, when the background affine connection is on-shell, the effective actions in the first and second order formulations coincide. While earlier works have verified this equivalence in a variety of ways, our goal here is to present a more direct and transparent derivation, confirmed by explicit one-loop calculations in a general background gauge.

The distinction between first- and second-order formulations of gravity is not merely a matter of formalism. It reflects different choices of fundamental variables, and potentially different paths to quantization. 
Establishing their quantum equivalence is also essential for confirming the internal consistency  of more general models, such as the Einstein–Cartan theory \cite{Hehl:1976kj, Shapiro:2001rz, Hehl:2023khc} or supergravity \cite{VanNieuwenhuizen:1981ae}, and of the background field approach.

In Sec. II, we derive a relation between the generating functionals of connected Green's functions in the first and second order formulations of the EH theory. From this, we obtain a basic identity between the corresponding background effective actions (see Eq.~\eqref{eq:216}), which explains in a simple and general way the equivalence observed when the background affine connection is on-shell. Based on this relation, we also argue that the counterterm Lagrangians in both formulations should coincide under the same condition. These features are explicitly verified at one-loop order in Sec. III, using a general background gauge. A brief discussion of the results is presented in Sec. IV .  Some useful details of the computations and additional results are given in the Appendices.

\section{Equivalence of formulations in the EH theory}
The Einstein-Hilbert Lagrangian in the second order formulation can be written in the form  \cite{Goldberg:1958zz}
\begin{equation}\label{eq:21}
    \mathcal{L}^{\text{II}} ( h) = -\frac{1}{\kappa^{2}} \sqrt{- g} g^{\mu \nu} R_{\mu \nu} ( \Gamma ) \equiv -\frac{1}{\kappa^{2}} h^{\mu \nu} R_{\mu \nu} ( \Gamma ) 
\end{equation}
where $\kappa^2 = 16\pi G_N$ ($G_{N} $ is Newton's gravitational constant), $g = \det g_{\mu\nu}$, the Goldberg metric $ h^{\mu \nu} = \sqrt{-g} g^{\mu \nu} $ and the Ricci tensor is given by \begin{equation} R_{\mu\nu} ( \Gamma )=  \Gamma_{\mu\lambda}{}^\lambda_{, \nu} -\Gamma_{\mu\nu }{}^\lambda_{, \lambda} - \Gamma_{\mu\nu}{}^\lambda \Gamma_{\lambda\sigma}{}^\sigma + \Gamma_{\mu\lambda}{}^\sigma \Gamma_{\nu\sigma}{}^\lambda,\end{equation}
where a comma denote a partial derivative.
The Levi-Civita connection is  expressed in terms of the metric as
\begin{equation}\Gamma_{\mu\nu}{}^\lambda = \frac{1}{2}g^{\lambda\rho}(g_{\mu\rho ,\nu} +g_{\nu\rho ,\mu} - g_{\mu\nu ,\rho}).
\end{equation}

In the background field method, one splits the gauge field $ h^{\mu \nu} $ as $ \bar{h}^{\mu \nu} + \kappa \mathfrak{h}^{\mu \nu}  $,  where $ \bar{h}^{\mu \nu} $ and $ \mathfrak{h}^{\mu \nu} $ denote respectively the background and quantum parts of this field.
Then, a term is added which breaks the gauge invariance of the quantum field. A convenient gauge-fixing Lagrangian $ \mathcal{L}_{\text{gf}} $, which is dependent upon the background field $ \bar{{h} }^{\mu \nu} $, may be chosen as
\begin{equation}\label{eq:22}
    \mathcal{L}_{\text{gf}} = - \frac{1}{2 \xi} \bar{h}_{\nu \beta} 
    (\bar{\mathsf{D}}_{\mu} \mathfrak{h}^{\mu \nu} )  
    (\bar{\mathsf{D}}_{\alpha} \mathfrak{h}^{\alpha \beta } )
\end{equation}
where $ \xi $ is a general gauge fixing parameter and $ \bar{\mathsf{D}} $  is the background covariant derivative. 
We  note that $ \mathcal{L}_{ \text{gf} } $ leads to a ghost Lagrangian induced by the Faddeev-Popov determinant 
\begin{equation}\label{eq:23}
    \begin{split}
        \Delta_{\text{FP}}(\bar{{h} } ) =
    \int \mathop{\mathcal{D} \bar{c}^{\mu}} \mathop{\mathcal{D} c^{\nu}} \exp -i \int \mathop{d^{{4}} x} 
    \sqrt{- \bar{h}} 
    \bar{h}_{\beta \nu} c^{\star \beta}
    \left[ 
        \bar{\mathsf{D}}_{\mu} 
        \left( \bar{h}^{\mu \lambda} \bar{\mathsf{D}}_{\lambda} c^{\nu} + \bar{h}^{\lambda \nu} \bar{\mathsf{D}}_{\lambda} c^{\mu} -    \bar{\mathsf{D}}_{\lambda} (\bar{h}^{\mu \nu} c^{\lambda} )\right)
\right]
    ,
\end{split}
\end{equation}
where $ \bar{h} \equiv \det \bar{h}^{\mu \nu} = {\bar{g}} $. 
The above Lagrangians are invariant under the gauge transformations 
    \begin{align}\label{eq:24}
        \Delta \bar{{h}}^{\mu \nu} 
        ={}&\bar{h}^{\lambda \nu} \partial_{\lambda} \zeta^{\mu} + \bar{h}^{\mu \lambda} \partial_{\lambda} \zeta^{\nu} - \partial_{\lambda} ( \bar{h}^{\mu \nu} \zeta^{\lambda} ), \quad 
        \Delta \mathfrak{h}^{\mu \nu} =\mathfrak{h}^{\lambda \nu} \partial_{\lambda} \zeta^{\mu} + \mathfrak{h}^{\mu \lambda} \partial_{\lambda} \zeta^{\nu} - \partial_{\lambda} ( \mathfrak{h}^{\mu \nu} \zeta^{\lambda} ); \quad 
\end{align}
where $ \zeta^{\mu} $ is an arbitrary  infinitesimal parameter. 

The first order Lagrangian has an identical form to that in Eq.~\eqref{eq:21}, but now $ h^{\mu \nu} $ and $ \tensor{\Gamma}{_{\mu \nu}^{\lambda}} $ are treated as independent fields. In this case, it is convenient to use, instead of $ \tensor{\Gamma}{_{\mu \nu}^{\lambda}} $, a related field $ \tensor{G}{_{\mu \nu}^{\lambda}} $ defined as 
\begin{equation} \label{eq:25a}
    \tensor{G}{_{\mu \nu}^{\lambda}} =
\tensor{\Gamma}{_{\mu \nu}^{\lambda } }-     \dfrac{1}{2} [ \delta_{\mu}^{\lambda} \tensor{\Gamma}{_{\nu \alpha }^{\alpha}} + \delta_{\nu}^{\lambda } \tensor{\Gamma}{_{\mu \alpha }^{\alpha}} ].
\end{equation}
In terms of this field, the corresponding first order Lagrangian may be written as \cite{Brandt:2016eaj}
\begin{equation}\label{eq:25}
    \mathcal{L}^{\text{I}} (h; G) = \frac{1}{ \kappa^{2}} \left(\frac{1}{2} \tensor{G}{_{\mu \nu}^{\lambda}} \tensor{M}{^{\mu \nu}_{\lambda}^{\alpha \beta}_{\gamma}} (h) \tensor{G}{_{\alpha \beta}^{\gamma}} - \tensor{G}{_{\mu \nu}^{\lambda }} \partial_{\lambda} h^{\mu \nu} \right),
\end{equation}
where, in ${{D}}$ spacetime dimensions,
\begin{equation}\label{eq:defM}
\begin{split}
        \tensor{M}{^{\mu\nu}_{\lambda}^{\pi\tau}_{\sigma}}(h)   &= 
        \frac{1}{2}\Big[\frac{1}{{{D}}-1}\left( \delta^\nu_\lambda\delta^\tau_\sigma h^{\mu\pi}+
                                                \delta^\mu_\lambda\delta^\tau_\sigma h^{\nu\pi}+
                                                \delta^\nu_\lambda\delta^\pi_\sigma h^{\mu\tau}+
                                                \delta^\mu_\lambda\delta^\pi_\sigma h^{\nu\tau}
\right) 
      \\  & \qquad  -  
\left( 
                                                \delta^\tau_\lambda\delta^\nu_\sigma h^{\mu\pi}+
                                                \delta^\tau_\lambda\delta^\mu_\sigma h^{\nu\pi}+
                                                \delta^\pi_\lambda\delta^\nu_\sigma h^{\mu\tau}+
                                                \delta^\pi_\lambda\delta^\mu_\sigma h^{\nu\tau}
                                        \right) \Big].
\end{split}
\end{equation}

We now  split the field  $ \tensor{G}{_{\mu \nu}^{\lambda}} $ into  a classical  background field $ \tensor{\bar{G}}{_{\mu \nu}^{\lambda} } $ and a  quantum field $ \tensor{\mathfrak{G}}{_{\mu \nu}^{\lambda}} $ as $\tensor{G}{_{\mu \nu}^{\lambda}} = \tensor{\bar{G}}{_{\mu \nu}^{\lambda}} + \kappa \tensor{\mathfrak{G}}{_{\mu \nu}^{\lambda}}$.
In this way, we obtain the following first order Lagrangian with background fields
\begin{equation}\label{eq:26}
    \mathcal{L}_{\text{BFM}}^{\text{I}}( \bar{h} , \mathfrak{h} ; \bar{G} , \mathfrak{G} )  
    = 
    \frac{1}{2}  \tensor{\mathfrak{G}}{_{\mu \nu}^{\lambda} }  \tensor{M}{^{\mu \nu}_{\lambda}^{\sigma \rho}_{\gamma}} ( h  )  \tensor{\mathfrak{G}}{_{\sigma \rho }^{\gamma} }  
    + 
     \tensor{\bar{G}}{_{\mu \nu}^{\lambda} }  \tensor{M}{^{\mu \nu}_{\lambda}^{\sigma \rho}_{\gamma}} ( \mathfrak{h}  )  \tensor{\mathfrak{G}}{_{\sigma \rho }^{\gamma} }  
    -  \tensor{\mathfrak{G}}{_{\mu \nu}^{\lambda}} \tensor{\mathfrak{h} }{^{\mu \nu}_{, \lambda}} 
 + \mathcal{L}^{\text{I} } ( \bar{h} ; \bar{G} ).
\end{equation}
This form was obtained by omitting the terms linear in the quantum fields, which can propagate only inside the loops. Such a procedure yields  the proper one-particle irreducible Green functions, when the background fields are off-shell \cite{Abbott:1980hw, weinberg:2000, Frenkel:2018xup}.

The Lagrangian \eqref{eq:26} is invariant under the infinitesimal background field transformations   
\begin{subequations}\label{eq:27}
    \begin{align}\label{eq:27a}
        \Delta \bar{h}^{\mu \nu} 
        ={}&\bar{h}^{\lambda \nu} \partial_{\lambda} \zeta^{\mu} + \bar{h}^{\mu \lambda} \partial_{\lambda} \zeta^{\nu} - \partial_{\lambda} ( \bar{h}^{\mu \nu} \zeta^{\lambda} ), \quad
        \Delta \mathfrak{h}^{\mu \nu} =\mathfrak{h}^{\lambda \nu} \partial_{\lambda} \zeta^{\mu} + \mathfrak{h}^{\mu \lambda} \partial_{\lambda} \zeta^{\nu} - \partial_{\lambda} ( \mathfrak{h}^{\mu \nu} \zeta^{\lambda} )
        \intertext{and}
        \nonumber
        \Delta \tensor{{\bar{G}}}{_{\mu \nu}^{\lambda}} ={}&  - \partial_{\mu} \partial_{\nu}  \zeta^{\lambda} + \frac{1}{2} ( \delta_{\mu}^{\lambda} \partial_{\nu} + \delta_{\nu}^{\lambda } \partial_{\mu} ) \partial_{\rho} \zeta^{\rho} - \zeta^{\rho} \partial_{\rho} \tensor{\bar{G}}{_{\mu \nu}^{\lambda}} + \tensor{\bar{G}}{_{\mu \nu}^{\rho}} \partial_{\rho} \zeta^{\lambda} -  ( \tensor{\bar{G}}{_{\mu \rho}^{\lambda}} \partial_{\nu} + \tensor{\bar{G}}{_{\nu \rho}^{\lambda}} \partial_{\mu}  ) \zeta^{\rho},
\\ 
        \Delta \tensor{{\mathfrak{G}}}{_{\mu \nu}^{\lambda}} ={}& 
        - \zeta^{\rho} \partial_{\rho} \tensor{\mathfrak{G}}{_{\mu \nu}^{\lambda}} + \tensor{\mathfrak{G}}{_{\mu \nu}^{\rho}} \partial_{\rho} \zeta^{\lambda} -  ( \tensor{\mathfrak{G}}{_{\mu \rho}^{\lambda}} \partial_{\nu} + \tensor{\mathfrak{G}}{_{\nu \rho}^{\lambda}} \partial_{\mu}  ) \zeta^{\rho}.
\end{align}
\end{subequations}

In order to compare the first and second order formulations, we note  that after functionally integrating over the quantum field $ \tensor{\mathfrak{G}}{_{\mu \nu}^{\lambda}}$ in \eqref{eq:26}, one gets up to irrelevant pure background terms, that 
\begin{equation}\label{eq:28}
    \begin{split}
        & \int \mathop{\mathcal{D} \tensor{\mathfrak{G}}{_{\mu \nu}^{\lambda}}} \exp i \int \mathop{d^{{4}} x} \left(\mathcal{L}^{\text{I}}_{\text{BFM}} - \mathcal{L}_{\text{BFM}}^{\text{II}} \right) \\ ={}& \exp i \int \mathop{d^{{4}} x} \bigg\{-  \left [ {\bar{G}} - {\mathcal{G}} ( \bar{h} )\right ]\tensor{}{_{\mu \nu}^{\lambda}}  \left (\frac{1}{2  }\tensor{ [M(\mathfrak{h}) M^{-1} ( h ) M(\mathfrak{h})]}{^{\mu \nu}_{\lambda}^{\alpha \beta}_{\gamma} } \tensor{[ \bar{G} + \mathcal{G} ( \bar{h} ) ]}{_{\alpha \beta }^{\gamma}} -  \tensor{[M( \mathfrak{h} ) M^{-1} ( h )]}{^{\mu \nu}_{\lambda}_{\alpha \beta}^{\gamma} } \partial_{\gamma} \mathfrak{h}^{\alpha \beta}\right )
        \bigg\}
        ,
    \end{split}
\end{equation}
where $ \tensor{\mathcal{G}}{_{\mu \nu}^{\lambda}} ( \bar{h} )$ is the on-shell value of the field $ \tensor{\bar{G}}{_{\mu \nu}^{\lambda}} $ following from Eq.~\eqref{eq:25}, which is given by
\begin{equation}\label{eq:29}
\left . \tensor{\bar{G}}{_{\mu \nu}^{\lambda}} \right|_{\text{on-shell}} = \tensor{\mathcal{G}}{_{\mu \nu}^{\lambda}} ( \bar{h} ) \equiv \tensor{(M^{-1})}{_{\mu \nu}^{\lambda}_{\alpha \beta}^{\gamma} } ( \bar{h} ) \partial_{\gamma} \bar{h}^{\alpha \beta}.
\end{equation}

The Eq.~\eqref{eq:28} shows that when the background field $ \bar{G} $ is on-shell, the first and the second order Lagrangians  in the background field method effectively become equivalent. Such a result indicates  that in this case, the effective actions  at one-loop order may also become equal.

This statement can be established by deriving basic relations between the generating functionals of Green's functions in the presence of background fields. To that end, we observe that, in the first order formulation, the functional takes the form
\begin{equation}\label{eq:210}
    \bar{Z}^{\text{I}} [ j,J; \bar{\mathfrak{h}},  \bar{G} ] = \int\mathop{\mathcal{D} \mathfrak{h}^{\mu \nu}} \mathop{\mathcal{D} \tensor{\mathfrak{G}}{_{\mu \nu}^{\lambda} } } \Delta_{\text{FP}} ( \bar{\mathfrak{h} }  ) \exp i \int \mathop{d^{{4}} x} \left [ \mathcal{L}^{\text{I}} (\bar{{h}} , \mathfrak{h}; \bar{G} , \mathfrak{G}  ) + \mathcal{L}_{\text{gf}}  + j_{\mu \nu} \mathfrak{h}^{\mu \nu} + \tensor{J}{^{\mu \nu}_{\lambda}} \tensor{\mathfrak{G}}{_{\mu \nu}^{\lambda} } \right ],  
\end{equation}
where we define $ \bar{h}^{\mu \nu} = \eta^{\mu \nu} + \kappa \bar{\mathfrak{h}}^{\mu \nu} $, $ j_{\mu \nu} $ and $ \tensor{J}{^{\mu \nu}_{\lambda}} $ are the sources of the quantum fields $ \mathfrak{h}^{\mu \nu} $ and $ \tensor{\mathfrak{G}}{_{\mu \nu}^{\lambda}} $. Similarly, we get in second order formulation 
\begin{equation}\label{eq:211}
    \bar{Z}^{\text{II}} [ j; \bar{\mathfrak{h}} ] = \int\mathop{\mathcal{D} \mathfrak{h}^{\mu \nu}}  \Delta_{\text{FP}} ( \bar{\mathfrak{h} }  ) \exp i \int \mathop{d^{{4}} x} \left [ \mathcal{L}^{\text{II} } (\bar{{h}} , \mathfrak{h} ) + \mathcal{L}_{\text{gf}}  + j_{\mu \nu} \mathfrak{h}^{\mu \nu} \right ].
\end{equation}
Using the Eqs.~\eqref{eq:28}--\eqref{eq:211}, setting $ \tensor{J}{^{\mu \nu}_{\lambda}} =0$ and the field $ \tensor{\bar{G}}{_{\mu \nu}^{\lambda}} $ on shell, we get the relation 
\begin{equation}\label{eq:212}
\bar{Z}^{\text{I}} [j,0; \bar{\mathfrak{h}} , \mathcal{G} ( \bar{\mathfrak{h}} )] =
    \bar{Z}^{\text{II} } [ j; \bar{\mathfrak{h}} ]. 
\end{equation}
A similar equation holds for the generating functionals of connected Green functions: $ \bar{W} =-i \ln{\bar{Z}} $.

Consider now the generating functionals of proper vertices $ \bar{\Gamma} $ which are related to $ \bar{W}  $ by the Legendre transformations
\begin{subequations}\label{eq:213and214}
    \begin{align}\label{eq:213}
        \bar{\Gamma}^{\text{I}} [ \hat{\mathfrak{h}} , \hat{\mathfrak{G}} ;\bar{\mathfrak{h}} , \bar{G}]
        ={}& \bar{W}^{\text{I}} [j,J; \bar{\mathfrak{h}} , \bar{G} ] - \int \mathop{d^{{4}}x} \left [ j_{\mu \nu} \hat{\mathfrak{h}}^{\mu \nu }  + \tensor{J}{^{\mu \nu}_{\lambda}} \tensor{\hat{\mathfrak{G}}}{_{\mu \nu}^{\lambda}}\right ], \\ \label{eq:214}
        \bar{\Gamma}^{\text{II}} [ \hat{\tilde{\mathfrak{h}}}  ;\bar{\mathfrak{h}} ]
        ={}&\bar{W}^{\text{II}} [j; \bar{\mathfrak{h}}  ] - \int \mathop{d^{{4}}x}  j_{\mu \nu} \hat{\tilde{\mathfrak{h}}}^{\mu \nu}  .
    \end{align}
\end{subequations}
Here, the mean quantum fields $ \hat{\tilde{\mathfrak{h}}}^{\mu \nu} $, $ \hat{\mathfrak{h}}^{\mu \nu} $ and $ \tensor{\hat{\mathfrak{G}}}{_{\mu \nu}^{\lambda}} $ are defined by 
\begin{equation}\label{eq:215}
    \hat{\tilde{\mathfrak{h}}}^{\mu \nu} = \frac{\delta \bar{W}^{\text{II}}}{\delta j_{\mu \nu}}, \quad 
    \hat{\mathfrak{h}}^{\mu \nu} = \frac{\delta \bar{W}^{\text{I}}}{\delta j_{\mu \nu}}; \quad \tensor{\hat{\mathfrak{G}}}{_{\mu \nu}^{\lambda}} = \frac{\delta \bar{W}^{\text{I}} }{\delta \tensor{J}{^{\mu \nu}_{\lambda}}}.
\end{equation}
Of special interest is the background effective action, obtained by evaluating $ \bar{\Gamma } $ for vanishing mean quantum fields. Thus, putting $J$ to zero, setting the  field $ \tensor{\bar{G}}{_{\mu \nu}^{\lambda}} $ on-shell and using Eqs.~\eqref{eq:29}, \eqref{eq:212} and Eq.~\eqref{eq:213and214}, one obtains  the basic relation 
\begin{equation}\label{eq:216}
    \bar{\Gamma}^{\text{I}} [ 0,0 ;\bar{\mathfrak{h}} , \mathcal{G} ( \bar{\mathfrak{{h}}} )]
        =
        \bar{\Gamma}^{\text{II}} [ 0 ;\bar{\mathfrak{h}} ].
\end{equation}
This equation directly  demonstrates the equivalence at the quantum level of the two formulations of the Einstein-Hilbert theory. As discussed in Appendix~\ref{sec:YM}, an analogous result also holds in Yang–Mills theory. In the following section, we verify Eq.~\eqref{eq:216} explicitly at one-loop order in a general background gauge.

\section{One-Loop Counterterms in a General Background Gauge} 

\begin{figure}[ht]
   \includegraphics[width=0.92\textwidth]{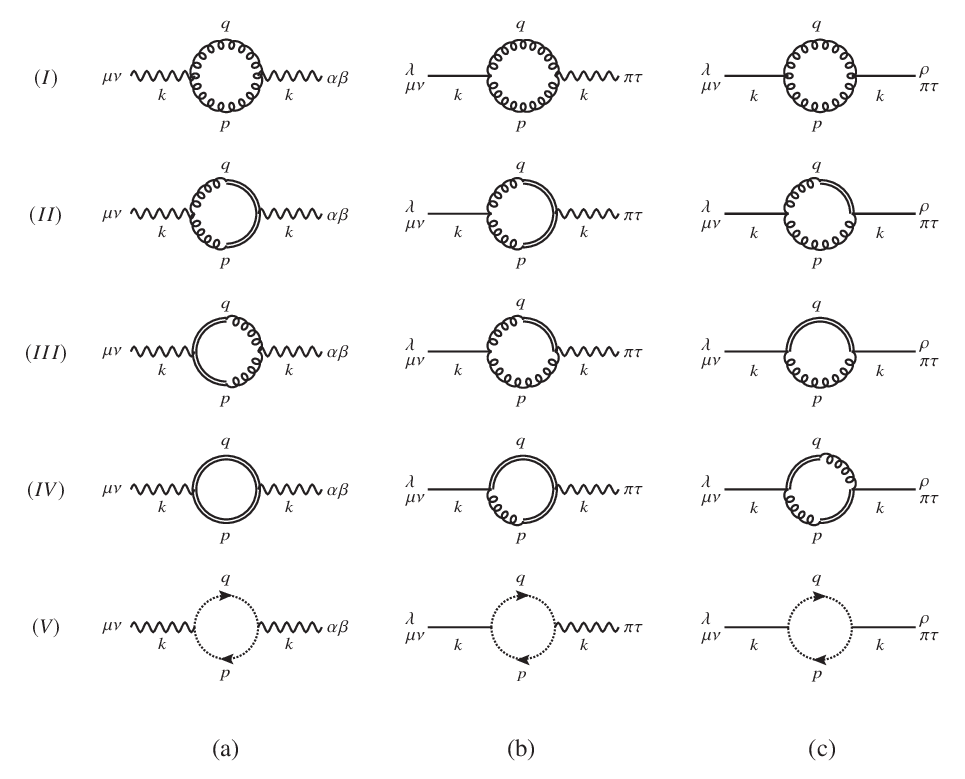}
    \caption{Diagrams that contributes to the self-energies $ \bar{\mathfrak{h}} \bar{\mathfrak{h}} $ (a), $ \bar{G} \bar{\mathfrak{h}} $ (b) and $ \bar{G} \bar{G} $ (c) in the first order formulation of the EH theory. Wavy and solid lines represents respectively the background fields $ \bar{\mathfrak{h} }$ and $ {\bar{G}}$. The quantum fields $ \mathfrak{h}$ and $ \mathfrak{G}$ are represented by springy and double solid lines. Momenta in the loops flow clockwise, so that, $ q = p+k$.}\label{fig:1}
\end{figure}

The one-loop diagrams contributing in the first order formulation to the background self-energies $ \bar{\mathfrak{h}} \bar{\mathfrak{h}} $, $ \bar{G} \bar{\mathfrak{h}} $ and $ \bar{G} \bar{G} $ are shown in Fig.~\ref{fig:1}, where we have defined: $ \bar{h}^{\mu \nu} = \eta^{\mu \nu} + \kappa \bar{\mathfrak{h}}^{\mu \nu} $. 

At one-loop order, the quadratic part of the effective Lagrangian can be written in terms of these self-energies as 
\begin{equation}\label{eq:31}
    \mathcal{L}^{\text{I}}_{(2)} ( \bar{\mathfrak{h}} , \bar{G} ) = \frac{1}{2} \left ( \bar{\mathfrak{h}}^{\mu \nu} \Pi_{\mu \nu \alpha \beta} \bar{\mathfrak{h}}^{\alpha \beta} + 2\tensor{\bar{G}}{_{\alpha \beta}^{\sigma}} \tensor{\Pi}{^{\alpha \beta}_{\sigma }_{\mu \nu} } \bar{\mathfrak{h}}^{\mu \nu} + \tensor{\bar{G}}{_{\mu \nu}^{\rho}} \tensor{\Pi}{^{\mu \nu}_{\rho}^{\alpha \beta}_{\sigma}} \tensor{\bar{G}}{_{\alpha \beta}^{\sigma}}\right ),
\end{equation}
where both background fields $ \bar{\mathfrak{h}}^{\mu \nu} $ and $ \tensor{\bar{G}}{_{\mu \nu}^{\lambda}} $ are off-shell. As shown in Eqs.~\eqref{eq:ID1} and \eqref{eq:ID2}, the above self-energies satisfy Ward identities that are induced by the background gauge invariance of the EH theory under the background field transformations~\eqref{eq:27}.
In a general background gauge, the Eq.~\eqref{eq:31} leads to a very involved expression depending upon the gauge parameter $ \xi $, which has been evaluated in Appendix~\ref{sec:OneLoop}.

Here, we discuss only the result obtained by setting the background field $ \tensor{\bar{G}}{_{\mu \nu}^{\lambda}} $ on-shell (see Eq.~\eqref{eq:29}), which is needed for the verification of Eq.~\eqref{eq:216}. In this case, Eq.~\eqref{eq:31} yields the proper background graviton self-energy in the first order formulation. 
This self-energy may be expressed in terms of five independent tensors built from $ \eta_{\mu \nu} $ and $ k_{\alpha} $, as 
\begin{subequations}\label{eq:32}
    \allowdisplaybreaks
    \begin{align}\label{eq:32a}
        \mathcal{T}^{(1)}{}_{\mu \nu \alpha \beta} (k) ={}&  \frac{k_{\mu} k_{\nu} k_{\alpha} k_{\beta}}{k^{4}}, \\
        \mathcal{T}^{(2)}{}_{\mu \nu \alpha \beta} (k) ={}& \eta_{\mu \nu} \eta_{\alpha \beta} , \\ 
        \mathcal{T}^{(3)}{}_{\mu \nu \alpha \beta} (k) ={}& \eta_{\mu \alpha} \eta_{\nu \beta } + \eta_{\mu \beta} \eta_{\nu \alpha},\\
    \mathcal{T}^{(4)}{}_{\mu \nu \alpha \beta} (k) ={}& \frac{1}{k^{2}}\left(k_{\mu} k_{\nu} \eta_{\alpha \beta} + k_{\alpha} k_{\beta} \eta_{\mu \nu}\right), \\
    \mathcal{T}^{(5)}{}_{\mu \nu \alpha \beta} (k) ={}& \frac{1}{k^{2} }\left(k_{\mu} k_{\alpha} \eta_{\nu \beta} + k_{\nu} k_{\alpha} \eta_{\mu \beta} + k_{\mu} k_{\beta} \eta_{\nu \alpha} + k_{\nu} k_{\beta} \eta_{\mu \alpha}\right). 
    \end{align}
\end{subequations}
Thus, we can write the divergent part of this self-energy in ${{D}}=4-2 \epsilon $ dimensions, in the form 
\begin{equation}\label{eq:33}
    \Pi^{\text{I div}}_{\mu \nu \alpha \beta} (k)= \frac{\kappa^{2} k^{4}}{16 \pi^{2} \epsilon} \sum_{i=1}^{5} C_{i} ( \xi ) \mathcal{T}^{(i)}{}_{\mu \nu \alpha \beta} (k), 
\end{equation}
where the coefficients $ C_{i} ( \xi )$ are dimensionless functions of the gauge parameter.

The proper background graviton self-energy satisfies the Ward identity
\begin{equation}\label{eq:34a}
( \eta^{\mu \rho} k^{\nu} + \eta^{\nu \rho} k^{\mu} - \eta^{\mu \nu} k^{\rho} ) 
    \Pi^{\text{I}}_{\mu \nu \alpha \beta} (k) =0,
\end{equation}
which is a consequence of gauge invariance under the background field transformations \eqref{eq:24}. The above identity implies that the coefficients $ C_{i} $ are not independent. Indeed, one finds after an explicit calculation, the relations 
\begin{subequations}\label{eq:35}
    \begin{align}\label{eq:35a}
        C_{1} ( \xi ) ={}& 4 \left[ C_{2} ( \xi ) + C_{3} (\xi )\right]  ,\\
        C_{2} ( \xi ) ={}& -\frac{20 \xi + 99}{240} 
        ,\\
        C_{3} ( \xi ) ={}&
        \frac{20 (\xi -1) \xi +87}{80}  
        ,\\
        C_{4} ( \xi ) ={}& 2 \left[ C_{2} ( \xi ) + C_{3} (\xi )\right],\\
        C_{5} ( \xi ) ={}& - C_{3} (\xi) .
    \end{align}
\end{subequations}

We can connect the self-energy \eqref{eq:33} to the counterterm Lagrangian by using the relations 
\begin{subequations}\label{eq:defRinh}
    \begin{align}\label{eq:36}
        \bar{R}_{\mu \nu} ={}& \frac{\kappa}{2} \left [ k_{\mu} k_{\nu} L_{\rho \sigma} - \frac{k^{2}}{2} \left ( L_{\mu \sigma } L_{\rho \nu} + L_{\mu \rho} L_{\sigma \nu}\right )\right ] \left ( \frac{1}{2} \eta^{\rho \sigma } \tensor{\bar{\mathfrak{h}}}{^{\tau}_{\tau} } - \bar{\mathfrak{h}}^{\rho \sigma }  \right ) + O ( \kappa^{2} ), \\
        \bar{R} ={}& \kappa k^{2} L_{\alpha \beta} \left ( \frac{1}{2} \eta^{\alpha \beta} \tensor{\bar{\mathfrak{h}}}{^{\sigma}_{\sigma} } - \bar{\mathfrak{h}}^{\alpha \beta}  \right ) + O ( \kappa^{2} ),
    \end{align}
\end{subequations}
where $ \bar{R}_{\mu \nu} $ is the Ricci tensor, $ \bar{R} $ is the curvature scalar and $ L_{\mu \nu} (k) = k_{\mu} k_{\nu} / k^{2} - \eta_{\mu \nu} $. In this way, we obtain for the graviton background counterterm Lagrangian, the result 
\begin{equation}\label{eq:38}
    \left .\mathcal{L}_{\text{CT}}^{\text{I}} \right |_{\bar{G} = \mathcal{G}} = 
        \frac{\sqrt{- \bar{g}} }{16 \pi^{2} \epsilon} \left [ -  \left ( 
                \frac{\xi - 1}{6}   + \frac{119}{120}
        \right ) \bar{R}^{2} + \left ( 
        \xi(\xi -1)  + \frac{87}{20}
\right ) \bar{R}_{\mu \nu} \bar{R}^{\mu \nu}\right ].
\end{equation}
For $\xi=1$, Eq.~\eqref{eq:38} reduces to the result obtained in Ref.~\cite{Goncalves:2017jxq} in the Goldberg parametrization (see Appendix~\ref{sec:Shapiro}), providing an explicit one-loop verification of the equivalence in Eq.~\eqref{eq:216}. 
This equivalence further implies that Eq.~\eqref{eq:38} yields the general counterterm Lagrangian of the second-order EH theory for arbitrary gauge parameter $\xi$.

\section{Discussion}
We have investigated the equivalence between the first and second order formulations of the Einstein–Hilbert theory in a general background gauge. As a key step, we derived Eq.~\eqref{eq:212}, which relates the generating functionals of connected Green’s functions in the two formulations. This relation leads directly to Eq.~\eqref{eq:216}, establishing that the corresponding background effective actions coincide at the quantum level when the background affine connection is on-shell. Notably, this equivalence remains valid even when the background graviton field is off-shell.
The closed form expression \eqref{eq:216} is a novel and gauge-invariant result, which directly demonstrates the equivalence of the background effective actions in the two formulations. 

The expression~\eqref{eq:216} has been verified through  explicit calculations of the divergent part of the background  metric and affine connections 
self-energies at one loop order. These functions have been computed in a general background gauge, with both background fields taken off-shell. We have shown that these self-energies satisfy Ward identities which reflect the diffeomorphism invariance of the EH theory. 
Such explicit calculations are useful as these may ascertain and clarify the formal arguments for the equivalence at the quantum level, of the first and second order formulations of the Einstein-Hilbert theory.

In some theories, the original fields may not be the optimal  variables to describe the physical content of the models. On the other hand, a change of variables could lead to enhanced physical insights. This feature also occurs in the first order formulation of the EH theory \cite{Brandt:2016eaj}. In this case, it is useful to make the change of variables 
\begin{equation}\label{eq:disc1}
    \tensor*{G}{*^\prime_{\mu \nu}^{\lambda}} 
    =     
    \tensor{G}{_{\mu \nu}^{\lambda}} + \tensor{(M^{-1})}{_{\mu \nu}^{\lambda}_{\alpha \beta}^{\sigma} } (h)\tensor{h}{^{\alpha \beta}_{,\sigma}},
\end{equation}
where $h^{\alpha \beta } $ is the graviton field, the affine connection  field $ \tensor{G}{_{\mu \nu}^{\lambda}} $ is defined in Eq.~\eqref{eq:25a} and the tensor $M$ is given in Eq.~\eqref{eq:defM}. 
Then, it turns out that the generating functional of Green's functions containing only external  graviton fields, directly reduces to the appropriate generating functional in the second order EH theory.

We point out here that the equivalence expressed in Eq.~\eqref{eq:216} is not a forthright consequence of the Kallosh–DeWitt theorem \cite{DeWitt:1964mxt, deWitt:1967ub, Kallosh:1974yh}, which guarantees the equality of S-matrix elements, as our analysis allows for an off-shell background graviton field. A natural extension of this work would be to consider the Einstein–Cartan theory, which is structurally more complex due to the presence of two gauge symmetries: diffeomorphism invariance and local Lorentz invariance \cite{Brandt:2024rsy, Brandt:2024kvs}. This generalization is currently being examined.

\begin{acknowledgments}
D.\ G.\ C\@. M\@. thanks Ann Aksoy for enlightening conversations. 
F.\ T.\ B\@.  thanks Gustavo A. Menezes for useful conversations.
F.\ T.\ B\@., J.\ F\@. thank CNPq (Brazil) for financial support. 
\end{acknowledgments}

\appendix

\section{Background self-energies in the first-order Einstein-Hilbert theory} \label{sec:OneLoop}
In this appendix, we will show the details for the computation of the divergent part of the self-energies $ \bar{\mathfrak{h}} \bar{\mathfrak{h}} $, $ \bar{G}  \bar{\mathfrak{h}}$ and $ \bar{G} \bar{G} $ of the first order EH theory.
In order to compute it, we adopt the Passarino–Veltman approach~\cite{Passarino:1978jh} (for instance, see Appendix B of Ref.~\cite{Brandt:2022und}), which systematically reduces tensorial loop integrals to scalar integrals. We will use dimensional regularization \cite{Leibbrandt:1975dj} with $ {{D}}= 4 - 2 \epsilon $, where $ \epsilon $ is taken to be a small parameter. 

In this method, the self-energies are expanded in terms of appropriate tensorial bases. Therefore, a suitable basis must be specified for each type of tensor structure that appears. In addition to the basis introduced in Eq.~\eqref{eq:32} for the graviton self-energy, we will also employ the following basis for tensor structures of the form:
\begin{itemize}
    \item  $ \langle G^{\lambda}_{\mu \nu}h^{\pi \tau }  \rangle$:
\begin{subequations}\label{eq:TensHp}
\allowdisplaybreaks
\begin{align}
    (\mathcal{T}^{Gh}_{1})^{\lambda}_{\mu\nu}{}^{\pi\tau}  & =   
 \dfrac{1}{4} (k^{\pi} \delta^{\lambda}_{\nu} \delta_{\mu}^{\tau}+k^{\pi} 
\delta^{\lambda}_{\mu}
   \delta_{\nu}^{\tau}+\delta^{\pi}_{ \nu} k^{\tau }
                                                      \delta^{\lambda}_{\mu}
                                                  +\delta^{\pi}_{ \mu } k^{\tau} \delta^{\lambda}_{\nu}), \label{eq:ao}
\\
(\mathcal{T}^{Gh}_2)^{\lambda}_{\mu\nu}{}^{\pi\tau}  & = 
 \dfrac{1}{2}  k^{\lambda } (\delta^{\pi}_{ \nu }
                                                    \delta_{\mu}^{\tau}+\delta^{\pi}_{
                                                    \mu } 
\delta_{\nu}^{ \tau }), 
\\
(\mathcal{T}^{Gh}_3)^{\lambda}_{\mu\nu}{}^{\pi\tau}  & =
k^{\lambda }  \eta^{\pi \tau } \eta_{\mu\nu}, 
\\
(\mathcal{T}^{Gh}_{4})^{\lambda}_{\mu\nu}{}^{\pi\tau}  & =
 \dfrac{1}{2}  \eta_{\mu\nu} (k^{\pi} \eta^{\lambda\tau}+\eta^{\lambda\pi} k^{\tau 
   }), 
   \\
(\mathcal{T}^{Gh}_5)^{\lambda}_{\mu\nu}{}^{\pi\tau}  & = 
 \dfrac{1}{4}  (\delta^{\pi}_{ \nu } k_{\mu}
                                                    \eta^{\lambda\tau}+\eta^{\lambda\pi} 
k_{\mu}
   \delta_{\nu}^{\tau}+\delta^{\pi}_{ \mu } k_{\nu}
                                                    \eta^{\lambda\tau}+\eta^{\lambda\pi} 
k_{\nu} \delta_{\mu}^{\tau }), 
\\
(\mathcal{T}^{Gh}_{6})^{\lambda}_{\mu\nu}{}^{\pi\tau}  & =
 \dfrac{1}{2}  \eta^{\pi \tau } (k_{\mu}
                                                      \delta^{\lambda}_{\nu}+k_{\nu} 
\delta^{\lambda}_{ \mu }), 
\\
(\mathcal{T}^{Gh}_7)^{\lambda}_{\mu\nu}{}^{\pi\tau}  & = 
 \dfrac{1}{2 k^2} k^{\pi} k^{\tau} (k_{\mu}
                                                    \delta^{\lambda}_{\nu}+k_{\nu} 
\delta^{\lambda}_{ \mu }),
\\
(\mathcal{T}^{Gh}_8)^{\lambda}_{\mu\nu}{}^{\pi\tau}  & =
\dfrac{1}{4 k^2} k^{\lambda } (k^{\pi} k_{\mu} \delta_{\nu}^{\tau}+k^{\pi} k_{\nu}
   \delta_{\mu}^{\tau}+\delta^{\pi}_{ \nu } k_{\mu} k^{\tau }+\delta^{\pi}_{ \mu
                                                  } k_{\nu} k^{\tau}), 
                                                  \\
(\mathcal{T}^{Gh}_9)^{\lambda}_{\mu\nu}{}^{\pi\tau}  & =
 \dfrac{1}{2 k^2} k_{\mu} k_{\nu} (k^{\pi} \eta^{\lambda\tau}+\eta^{\lambda\pi} 
   k^{\tau }), 
   \\
(\mathcal{T}^{Gh}_{10})^{\lambda}_{\mu\nu}{}^{\pi\tau}  & =
\dfrac{1}{k^4}
k^{\lambda }   k_{\mu} k_{\nu}  k^{\pi}   k^{\tau}, 
\\
(\mathcal{T}^{Gh}_{11})^{\lambda}_{\mu\nu}{}^{\pi\tau}  & =
 \dfrac{1}{k^2}  k^{\lambda } \eta_{\mu\nu}  k^{\pi} k^{\tau}, \\
(\mathcal{T}^{Gh}_{12})^{\lambda}_{\mu\nu}{}^{\pi\tau}  & = 
\dfrac{1}{k^2} k^{\lambda } k_{\mu} k_{\nu} \eta^{\pi \tau }.
\end{align}
\end{subequations}

\item {$ \langle G_{\mu \nu}^{\lambda }  G_{\pi \tau}^{\rho} \rangle $}:

\begin{subequations}\label{eq:TensGG}
\allowdisplaybreaks
\begin{align}
 (\mathcal{T}^{GG}_{1}) ^{\lambda}_{\mu\nu}{}^\rho_{\pi\tau}
  & = \dfrac{1}{4}  (\delta_\pi^{\rho } \delta^{\lambda}_{\nu} \eta_{\mu\tau}+\delta_\pi^{\rho }
   \delta^{\lambda}_{\mu} \eta_{\nu\tau}+\eta_{\pi \nu } \delta^{\lambda}_{\mu} \delta^\rho_{ \tau }+\eta_{\pi \mu 
   } \delta^{\lambda}_{\nu} \delta^\rho_{ \tau }), 
   \\
 (\mathcal{T}^{GG}_{2})^{\lambda}_{\mu\nu}{}^\rho_{\pi\tau}  & = \dfrac{1}{2}  \eta^{\lambda \rho } (\eta_{\pi\nu } \eta_{\mu\tau}+\eta_{\pi \mu } \eta_{\nu \tau }), 
 \\
 (\mathcal{T}^{GG}_{3})^{\lambda}_{\mu\nu}{}^\rho_{\pi\tau}  & = \eta_{\pi \tau 
                                                       } \eta^{\lambda \rho }
                                                       \eta_{\mu\nu}, 
                                                       \\
 (\mathcal{T}^{GG}_{4})^{\lambda}_{\mu\nu}{}^\rho_{\pi\tau}  & = \dfrac{1}{4}  (\eta_{\pi\nu } \delta^{\lambda}_{\tau} \delta_\mu^{\rho }+\delta_{\pi}^{ \lambda }
   \delta_\mu^{\rho } \eta_{\nu \tau}+\eta_{\pi \mu } \delta^{\lambda}_{\tau} \delta_\nu^{\rho }+\delta_{\pi}^{ \lambda } \eta_{\mu\tau} \delta_\nu^{\rho }), 
   \\
 (\mathcal{T}^{GG}_{5})^{\lambda}_{\mu\nu}{}^\rho_{\pi\tau}  & = \dfrac{1}{4}  (\delta_\pi^{\rho } \delta^{\lambda}_{\tau} \eta_{\mu\nu}+\delta_{\pi}^{ \lambda }
   \eta_{\mu\nu} \delta^\rho_{ \tau }+\eta_{\pi \tau } \delta^{\lambda}_{\nu} \delta_\mu^{\rho }+\eta_{\pi \tau }
   \delta^{\lambda}_{\mu} \delta_\nu^{\rho }), 
   \\
 (\mathcal{T}^{GG}_{6})^{\lambda}_{\mu\nu}{}^\rho_{\pi\tau}  & = \dfrac{1}{4 k^4}  (k_{\pi } k^{\lambda } k_{\mu} k_{\nu } \delta^\rho_{ \tau }+\delta_\pi^{\rho 
   } k^{\lambda } k_{\mu} k_{\nu } k_{\tau }+k_{\pi } k_{\mu} k^{\rho } k_{\tau }
   \delta^{\lambda}_{\nu}+k_{\pi } k_{\nu } k^{\rho } k_{\tau } \delta^{\lambda}_{\mu}), 
   \\
(\mathcal{T}^{GG}_{7})^{\lambda}_{\mu\nu}{}^\rho_{\pi\tau}  & = \dfrac{1}{4 k^4}  k^{\lambda } k^{\rho } (k_{\pi } k_{\mu} \eta_{\nu \tau}+k_{\pi }
   k_{\nu } \eta_{\mu\tau}+\eta_{\pi  \nu } k_{\mu} k_{\tau }+\eta_{\pi \mu } k_{\nu } k_{\tau 
   }), 
   \\
 (\mathcal{T}^{GG}_{8})^{\lambda}_{\mu\nu}{}^\rho_{\pi\tau}  & = \dfrac{1}{4 k^4} 
                                                       (k_{\pi }
                                                       k_{\mu} k_{\nu 
                                                       } k^{\rho }
                                                       \delta^{\lambda}_{
                                                       \tau 
                                                       }+\delta_{\pi}^{ \lambda } k_{\mu} k_{\nu } k^{\rho } k_{\tau }+k_{\pi } k^{\lambda } k_{\mu} k_{\tau 
   } \delta_\nu^{\rho }+k_{\pi } k^{\lambda } k_{\nu } k_{\tau } \delta_\mu^{\rho }), 
   \\
 (\mathcal{T}^{GG}_{9})^{\lambda}_{\mu\nu}{}^\rho_{\pi\tau}  & = \dfrac{1}{2 k^4}  k^{\lambda } k^{\rho } (\eta_{\pi \tau } k_{\mu} k_{\nu }+k_{\pi }
   k_{\tau } \eta_{\mu\nu}), 
   \\
 (\mathcal{T}^{GG}_{10})^{\lambda}_{\mu\nu}{}^\rho_{\pi\tau}  & = \dfrac{1}{4 k^2}  (k_{\pi } \delta^\rho_{ \tau }+\delta_\pi^{\rho } k_{\tau }) (k_{\mu 
   } \delta^{\lambda}_{\nu}+k_{\nu } \delta^{\lambda}_{\mu}), 
   \\
 (\mathcal{T}^{GG}_{11})^{\lambda}_{\mu\nu}{}^\rho_{\pi\tau}  & = \dfrac{1}{4 k^2}  \eta^{\lambda \rho } (k_{\pi } k_{\mu} \eta_{\nu \tau}+k_{\pi } k_{\nu }
   \eta_{\mu\tau}+\eta_{\pi  \nu } k_{\mu} k_{\tau }+\eta_{\pi \mu } k_{\nu } k_{\tau }), 
   \\
 (\mathcal{T}^{GG}_{12})^{\lambda}_{\mu\nu}{}^\rho_{\pi\tau}  & = \dfrac{1}{4 k^2}  (k_{\pi } \delta^{\lambda}_{\tau}+\delta_{\pi}^{ \lambda } k_{\tau }) 
   (k_{\mu} \delta_\nu^{\rho }+k_{\nu } \delta_\mu^{\rho }), 
   \\
 (\mathcal{T}^{GG}_{13})^{\lambda}_{\mu\nu}{}^\rho_{\pi\tau}  & = \dfrac{1}{4 k^2}  (\delta_\pi^{\rho } k_{\mu} k_{\nu } \delta^{\lambda}_{\tau}+\delta_{\pi}^{ \lambda }
   k_{\mu} k_{\nu } \delta^\rho_{ \tau }+k_{\pi } k_{\tau }
                                                          \delta^{\lambda}_{\nu} 
\delta_\mu^{\rho }+k_{\pi } k_{\tau } \delta^{\lambda}_{\mu} \delta_\nu^{\rho }), 
\\
 (\mathcal{T}^{GG}_{14})^{\lambda}_{\mu\nu}{}^\rho_{\pi\tau}  & =  \dfrac{1}{2 k^2}  \eta^{\lambda \rho } (\eta_{\pi \tau } k_{\mu} k_{\nu }+k_{\pi } k_{\tau 
   } \eta_{\mu\nu}), 
   \\\nonumber
 (\mathcal{T}^{GG}_{15})^{\lambda}_{\mu\nu}{}^\rho_{\pi\tau}  & =  \dfrac{1}{8 k^2}  (\delta_\pi^{\rho } k^{\lambda } k_{\mu} \eta_{\nu \tau}+k_{\pi } k^{\rho 
   } \delta^{\lambda}_{\mu} \eta_{\nu \tau}+\delta_\pi^{\rho }
                                                          k^{\lambda }
                                                          k_{\nu } 
\eta_{\mu \tau 
   }+\eta_{\pi \nu } k^{\lambda } k_{\mu} \delta^\rho_{ \tau } \\ 
& \quad \qquad +  \eta_{\pi \mu } k^{\lambda } k_{\nu }
   \delta^\rho_{ \tau }+k_{\pi } k^{\rho } \delta^{\lambda}_{\nu}
    \eta_{\mu\tau}+
\eta_{\pi \nu } k^{\rho }
   k_{\tau } \delta^{\lambda}_{\mu}+\eta_{\pi \mu } k^{\rho } k_{\tau } \delta^{\lambda}_{\nu}), 
   \\
 (\mathcal{T}^{GG}_{16})^{\lambda}_{\mu\nu}{}^\rho_{\pi\tau}  & = \dfrac{1}{2 k^2}  k^{\lambda } k^{\rho } (\eta_{\pi \nu } \eta_{\mu\tau}+\eta_{\pi \mu }
   \eta_{\nu \tau}), 
   \\
 (\mathcal{T}^{GG}_{17})^{\lambda}_{\mu\nu}{}^\rho_{\pi\tau}  & = \dfrac{1}{4 k^2}  (\eta_{\pi \tau } k^{\lambda } k_{\mu} \delta_\nu^{\rho }+\eta_{\pi \tau }
   k^{\lambda } k_{\nu } \delta_\mu^{\rho }+k_{\pi } k^{\rho } \delta^{\lambda}_{\tau} \eta_{\mu \nu 
   }+\delta_{\pi}^{ \lambda } k^{\rho } k_{\tau } \eta_{\mu\nu}), 
   \\\nonumber
 (\mathcal{T}^{GG}_{18})^{\lambda}_{\mu\nu}{}^\rho_{\pi\tau}  & = \dfrac{1}{8 k^2}  (k_{\pi } k^{\lambda } \eta_{\mu\tau} \delta_\nu^{\rho }+\eta_{\pi \mu }
   k^{\lambda } k_{\tau } \delta_\nu^{\rho }+k_{\pi } k^{\lambda } \delta_\mu^{\rho } \eta_{\nu \tau 
   }+\eta_{\pi \nu } k_{\mu} k^{\rho } \delta^{\lambda}_{\tau} 
\\
&\quad \qquad+ \delta_{\pi}^{ \lambda } k_{\mu} k^{\rho }
   \eta_{\nu \tau}+\eta_{\pi \mu } k_{\nu } k^{\rho } \delta^{\lambda}_{\tau}+\delta_{\pi}^{\lambda } k_{\nu 
   } k^{\rho } \eta_{\mu\tau}+\eta_{\pi \nu } k^{\lambda } k_{\tau } \delta_\mu^{\rho }), 
   \\
 (\mathcal{T}^{GG}_{19})^{\lambda}_{\mu\nu}{}^\rho_{\pi\tau}  & = \dfrac{1}{4 k^2}  (k_{\pi } k^{\lambda } \eta_{\mu\nu} \delta^\rho_{ \tau }+\eta_{\pi \tau }
   k_{\mu} k^{\rho } \delta^{\lambda}_{\nu}+\eta_{\pi \tau } k_{\nu }
                                                          k^{\rho } 
\delta^{\lambda}_{ \mu 
   }+\delta_\pi^{\rho } k^{\lambda } k_{\tau } \eta_{\mu\nu}), 
   \\
(\mathcal{T}^{GG}_{20})^{\lambda}_{\mu\nu}{}^\rho_{\pi\tau}  & = 
\dfrac{1}{k^6} k_{\pi } k^{\lambda } k_{\mu} k_{\nu } k^{\rho }
                                                      k_{\tau }, 
                                                      \\
 (\mathcal{T}^{GG}_{21})^{\lambda}_{\mu\nu}{}^\rho_{\pi\tau}  & = \dfrac{1}{k^4}
                                                       k_{\pi } k_{\mu 
                                                       } k_{\nu }
                                                       k_{\tau }
                                                       \eta^{\lambda 
                                                       \rho }, 
                                                       \\
 (\mathcal{T}^{GG}_{22})^{\lambda}_{\mu\nu}{}^\rho_{\pi\tau}  & = 
                                                       \dfrac{1}{k^2}\eta_{\pi \tau } k^{\lambda } k^{\rho } \eta_{\mu\nu} . 
\end{align}
\end{subequations}
\end{itemize}

\subsection{Feynman rules}
The Feynman rules can be obtained from the BFM effective action:
\begin{equation}\label{eq:aa2}
    \frac{1}{2}  \tensor{\mathfrak{G}}{_{\mu \nu}^{\lambda} }  \tensor{M}{^{\mu \nu}_{\lambda}^{\sigma \rho}_{\gamma}} ( \bar{h}  )  \tensor{\mathfrak{G}}{_{\sigma \rho }^{\gamma} }  
    + 
    \tensor{\bar{G}}{_{\mu \nu}^{\lambda} }  \tensor{M}{^{\mu \nu}_{\lambda}^{\sigma \rho}_{\gamma}} ( \mathfrak{h}  )  \tensor{\mathfrak{G}}{_{\sigma \rho }^{\gamma} }  
    - \tensor{\mathfrak{G}}{_{\mu \nu}^{\lambda}} \tensor{\mathfrak{h} }{^{\mu \nu}_{, \lambda}} 
     + \mathfrak{L}_{\text{gf}} (h) + \mathfrak{L}_{\text{gh}} (h),
\end{equation}
where 
\begin{equation}\label{eq:aa4}
    \mathfrak{L}_{\text{gf}} (h) = - \frac{1}{2 \xi} \bar{h}_{\nu \beta}  
    \left[\bar{\mathsf{D}}_{\mu} \mathfrak{h}^{\mu \nu}  \right]  
    \left[\bar{\mathsf{D}}_{\alpha} \mathfrak{h}^{\alpha \beta }   \right]
\end{equation}
and 
\begin{equation}\label{eq:aa5}
    \mathfrak{L}_{\text{gh}} (h) = 
    -\sqrt{-\bar{h}} 
    \bar{h}_{\beta \nu} c^{\star \beta}\left[ \bar{\mathsf{D}}_{\mu} 
    \left( \bar{h}^{\mu \lambda} \bar{\mathsf{D}}_{\lambda} c^{\nu} + \bar{h}^{\lambda \nu} \bar{\mathsf{D}}_{\lambda} c^{\mu} -    \bar{\mathsf{D}}_{\lambda} (\bar{h}^{\mu \nu} c^{\lambda} )\right)\right].
\end{equation}

Eq.~\eqref{eq:aa5} follows from the BRST form of the ghost action:
\begin{equation}\label{eq:ghostBRST}
    -\sqrt{-\bar{h}} 
    \bar{h}_{\beta \nu} d^{\star \beta}\left[ \bar{\mathsf{D}}_{\mu} \left( \bar{h}^{\mu \lambda} \partial_{\lambda} d^{\nu} + \bar{h}^{\lambda \nu} \partial_{\lambda} d^{\mu} -    \partial_{\lambda} (\bar{h}^{\mu \nu} d^{\lambda} )\right)\right]
    .
\end{equation}
It can be written in a more familiar form as
\begin{equation}\label{eq:p1}
    -\sqrt{-\bar{h}}   
    c^{\star \mu}\left[  \bar{h}_{\mu \nu} \bar{h}^{\gamma \lambda} \bar{\mathsf{D}}_{\lambda} \bar{\mathsf{D}}_{\gamma}  - \bar{R}_{\nu \mu} ( \bar{G} ) + \mathsf{X}_{\mu \nu} \right] c^{\nu},
\end{equation}
where 
\begin{equation}\label{eq:defRG}
    R_{\nu \mu} =
-{G}_{ \mu \nu}{}^{ \lambda}{}_{, \lambda}  - \frac{1}{{{D}}-1} \left({G}_{ \mu \lambda}{}^{\lambda} {G}_{ \nu \sigma}{}^{ \sigma} + \partial_{\nu} \tensor{G}{_{\mu \sigma}^{\sigma}} - \partial_{\mu} \tensor{G}{_{\nu \sigma}^{\sigma}} \right) + {G}_{ \mu \sigma}{}^{ \lambda}{G}_{ \nu \lambda}{}^{ \sigma}  
\end{equation}
and 
\begin{equation}\label{eq:defX}
    \mathsf{X}_{\mu \nu} = 
    \bar{g}_{\mu \nu} \bar{\mathsf{D}}_{\rho} \bar{h}^{\rho \lambda} \bar{\mathsf{D}}_{\lambda}   - \bar{g}_{\mu \lambda} (\bar{\mathsf{D}}_{\rho} \bar{\mathsf{D}}_{\nu} \bar{h}^{\rho \lambda}) \mathds{1}- \bar{g}_{\mu \lambda} \bar{\mathsf{D}}_{\rho} \bar{h}^{\rho \lambda} \bar{\mathsf{D}}_{\nu} . 
\end{equation}
When metricity $ D_{\mu} g_{\alpha \beta} = 0$ is assumed, $ \mathsf{X}_{\mu \nu} = 0$ and $ R_{\nu \mu} = R_{\mu \nu} $ and Eq.~\eqref{eq:p1} reduces to the form found in 't Hooft and Veltman \cite{tHooft:1974toh}.

To obtain Eq.~\eqref{eq:p1}, we used that
\begin{equation}\label{eq:aa5inv}
    \bar{h}_{\beta \nu} c^{\star \beta}\left[ \bar{\mathsf{D}}_{\mu} 
    \left( \bar{h}^{\mu \lambda} \bar{\mathsf{D}}_{\lambda} c^{\nu} + \bar{h}^{\lambda \nu} \bar{\mathsf{D}}_{\lambda} c^{\mu} -    \bar{\mathsf{D}}_{\lambda} (\bar{h}^{\mu \nu} c^{\lambda} )\right)\right]=
    c^{\star \mu}
    \left(  \bar{h}_{\mu \nu} \bar{h}^{\gamma \lambda} \bar{\mathsf{D}}_{\lambda} \bar{\mathsf{D}}_{\gamma}    + [ \bar{\mathsf{D}}_{\nu} , \bar{\mathsf{D}}_{\mu}  ] + \mathsf{X}_{\mu \nu}  \right) c^{\nu}
\end{equation}
and
\begin{equation}\label{eq:p3}
    [ \mathsf{D}_{\nu} , \mathsf{D}_{\mu} ] c^{\nu}  = -R {}^{\nu}{}_{\beta}{}_{\nu \mu }  c^{\beta }=   -R_{ \nu\mu } c^{\nu}.
\end{equation}

The covariant derivative of a tensor in the first order formulation
\begin{equation}\label{eq:devDin2st}
    \mathsf{D}_{\mu} \tensor{T}{_{\alpha }^{\beta}} = \partial_{\mu} \tensor{T}{_{\alpha}^{\beta}} + \tensor{\Gamma}{_{\mu}_{\gamma}}^{\beta} \tensor{T}{_{\alpha}^{\gamma}}  
- \tensor{\Gamma}{_{\mu}_{\alpha}}^{\gamma} \tensor{T}{_{\gamma}^{\beta}}.
\end{equation}
Using the field $ \tensor{G}{_{\mu \nu}^{\lambda}} $ instead of the affine connection $ \tensor{\Gamma}{_{\mu \nu}^{\lambda}} $ yields
\begin{equation}\label{eq:devDin1st}
    \mathsf{D}_{\mu} \tensor{T}{_{\alpha}^{\beta}} = \partial_{\mu} \tensor{T}{_{\alpha}^{\beta}} + \tensor{G}{_{\mu}_{\gamma}}^{\beta} \tensor{T}{_{\alpha}^{\gamma}}  
- \tensor{G}{_{\mu}_{\alpha}}^{\gamma} \tensor{T}{_{\gamma}^{\beta}}
+ \frac{1}{1 - {{D}}} \left (
        \delta^\beta_\mu G_{\sigma\gamma}{}^\sigma + \delta^\beta_\nu G_{\sigma\gamma}{}^\sigma 
        -\delta^\gamma_\mu G_{\sigma\alpha}{}^\sigma - \delta^\gamma_\alpha G_{\sigma\mu}{}^\sigma 
\right ) 
,
\end{equation}
where we used the inverse of Eq.~\eqref{eq:25a}:
\begin{equation}\label{eq:25inverse}
    \Gamma_{\mu\nu}{}^\lambda = G_{\mu\nu}{}^\lambda + \frac{1}{1-{{D}}} \left( 
        \delta^\lambda_\mu G_{\sigma\nu}{}^\sigma + \delta^\lambda_\nu G_{\sigma\mu}{}^\sigma 
\right).
\end{equation}


\subsubsection{Propagators}
The propagators are the same as those found in the conventional formulation (see Ref.~\cite{Brandt:2016eaj}). They are obtained from the inverse of the bilinear terms in the quantum fields in Eq.~\eqref{eq:aa2}: 
\begin{equation}\label{eq:matrix}
     \begin{bmatrix}
        \boldsymbol{A} & \boldsymbol{B} \\
        \boldsymbol{C} & \boldsymbol{D}
        \end{bmatrix}
        \equiv  \begin{bmatrix}
        -\frac{1}{4 \xi} \left (
k^{\tau} \left(k^{\rho}
   \eta^{\pi\sigma}+k^{\sigma}
   \eta^{\pi\rho}\right) +  k^{\pi}
   \left(k^{\rho} \eta^{\tau\sigma}+k^{\sigma}
   \eta^{\rho\tau}\right)
        \right ) 
        & 
-ik^{\gamma } \left(\eta^{\pi \beta }
   \eta^{\alpha \tau}+\eta^{\pi \alpha }
   \eta^{\tau \beta }\right)
        \\
ik^{\lambda } \left(\eta^{\mu \sigma }
   \eta^{\rho  \nu }+\eta^{\mu \rho }
   \eta^{\nu \sigma }\right)
                                                      & \tensor{M}{_{\mu \nu}^{\lambda}_{\alpha \beta}^{\gamma}} ( \eta )  
    \end{bmatrix},
\end{equation}
in momentum space $ i \partial_{\mu} = k_{\mu} $. The inverse can be computed using the expression: 
\begin{equation}\label{eq:invB}
      \begin{bmatrix}
        \boldsymbol{A} & \boldsymbol{B} \\
        \boldsymbol{C} & \boldsymbol{D}
    \end{bmatrix}^{-1}
    = \begin{pmatrix}
        \boldsymbol{X}^{-1} & - \boldsymbol{X}^{-1} \boldsymbol{B} \boldsymbol{D}^{-1} \\
        - \boldsymbol{D}^{-1} \boldsymbol{C} \boldsymbol{X}^{-1} & \boldsymbol{D}^{-1} + \boldsymbol{D}^{-1} \boldsymbol{C} \boldsymbol{X}^{-1} \boldsymbol{B} \boldsymbol{D}^{-1}
    \end{pmatrix},
\end{equation}
where $ \boldsymbol{X} = \boldsymbol{A} - \boldsymbol{B} \boldsymbol{D}^{-1} \boldsymbol{C}$ is the Schur complement. The inverse of $ \boldsymbol{D} $ is given by:
\begin{equation}\label{eq:28a}
  \begin{split}
    \tensor{(M^{-1})}{_{\mu \nu}^{\lambda}_{ \pi \tau}^{ \rho}} (h) 
={}&- \frac{1}{2({{D}}-2)} h^{\lambda \rho} h_{\mu \nu} h_{\pi \tau} +
    \frac{1}{4} h^{\lambda \rho} \left ( h_{\pi \mu} h_{\tau \nu} + h_{\pi \nu} h_{\tau \mu}\right ) 
    \\
    & - \frac{1}{4} \left ( h_{\tau \mu} \delta_{\nu}^{\rho} \delta_{\pi}^{\lambda} + h_{\pi \mu} \delta_{\nu}^{\rho} \delta_{\tau}^{\lambda} +  h_{\tau \nu} \delta_{\mu}^{\rho} \delta_{\pi}^{\lambda} +  h_{\pi \nu} \delta_{\mu}^{\rho} \delta_{\tau}^{\lambda}\right ).
\end{split}
\end{equation}

For reference, the propagators used to compute the self-energy in this work are given by:  
\begin{subequations}\label{eq:FREH1dpr}
\begin{align}
    \vcenter{\hbox{\includegraphics[scale=0.6]{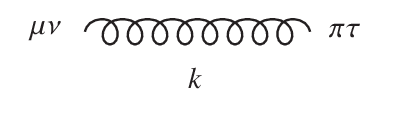}}} 
    & 
    \quad \quad 
        \mathcal{D }^{\mu \nu \pi \tau} (k),
    \\
    \vcenter{\hbox{\includegraphics[scale=0.6]{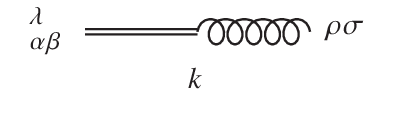}}} 
    & 
    \quad \quad 
   \tensor{\mathcal{D}}{_{\alpha \beta}^{\lambda}^{\rho \sigma}}(k),
\\
    \vcenter{\hbox{\includegraphics[scale=0.6]{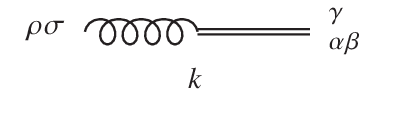}}} 
    & 
    \quad \quad
\tensor{\mathcal{D}}{^{\rho \sigma}_{\alpha \beta}^{\lambda}}(k),
    \\
    \vcenter{\hbox{\includegraphics[scale=0.6]{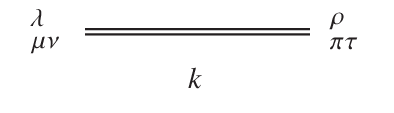}}} 
    & 
    \quad \quad 
\tensor{\mathcal{D}}{_{\mu \nu}^{\lambda}_{\pi \tau}^{\rho} }(k);
\end{align}
\end{subequations}
where
\begin{subequations}\label{eq:props}
    \begin{alignat}{3}\label{eq:hh}\nonumber
        & \mathcal{D }^{\mu \nu \pi \tau} (k) 
        ={}&&
        - \frac{i}{k^2} \big[\eta^{\pi \nu } \eta^{\mu \tau }+\eta^{\pi \mu } \eta^{\nu
 \tau } -(2- \xi )\eta^{\pi \tau } \eta^{\mu \nu }
+(\xi -1 )( k^{\pi } k^{\mu } \eta^{\nu \tau }+k^{\pi
    } k^{\nu } \eta^{\mu \tau } \\ & && \qquad + k^{\mu } k^{\tau }\eta^{\pi \nu }+ k^{\nu }k^{\tau }\eta^{\pi \mu }
-2\eta^{\pi \tau } k^{\mu } k^{\nu }-2k^{\pi } k^{\tau } \eta^{\mu \nu})\big],
   \\\nonumber 
   & \tensor{\mathcal{D}}{_{\alpha \beta}^{\lambda}^{\rho \sigma}}(k) 
    ={}&&
   \frac{1}{k^{2}} \Big[       (\mathcal{T}^{Gh}_2){}^{\lambda}_{\alpha \beta }{}^{\rho \sigma }  
    - 2  (\mathcal{T}^{Gh}_5){}^{\lambda}_{\alpha \beta }{}^{\rho \sigma} - (\xi -2)  (\mathcal{T}^{Gh}_6){}^{\lambda}_{\alpha \beta }{}^{\rho \sigma} \\ & && \qquad 
+( \xi  -1)  \left(2(\mathcal{T}^{Gh}_7){}^{\lambda}_{\alpha \beta }{}^{\rho \sigma} -2   (\mathcal{T}^{Gh}_9){}^{\lambda}_{\alpha \beta }{}^{\rho \sigma} +  (\mathcal{T}^{Gh}_{12}){}^{\lambda}_{\alpha \beta }{}^{\rho \sigma} \right)\Big],\\
                                                                                                                                                                 & \tensor{\mathcal{D}}{^{\rho \sigma}_{\alpha \beta}^{\lambda}}(k) 
={}&& -
\tensor{\mathcal{D} }{_{\alpha \beta}^{\lambda}^{\rho \sigma}}(k)
,\\\nonumber
   & \tensor{\mathcal{D}}{_{\mu \nu}^{\lambda}_{\pi \tau}^{\rho} }(k)
={}&&
\frac{i}{4}  \bigg[
    2(\mathcal{T}^{GG}_{2}){}^{\lambda}_{\mu\nu}{}^\rho_{\pi\tau}  
    - \frac{2}{{{D}}-2} (\mathcal{T}^{GG}_{3}){}^{\lambda}_{\mu\nu}{}^\rho_{\pi\tau}  
    -4(\mathcal{T}^{GG}_{4}){}^{\lambda}_{\mu\nu}{}^\rho_{\pi\tau}  
    +  8( \xi -1)(\mathcal{T}^{GG}_{6}){}^{\lambda}_{\mu\nu}{}^\rho_{\pi\tau}  
\\\nonumber & && \quad + 4( \xi -2)(\mathcal{T}^{GG}_{10}){}^{\lambda}_{\mu\nu}{}^\rho_{\pi\tau}  
-4(\mathcal{T}^{GG}_{11}){}^{\lambda}_{\mu\nu}{}^\rho_{\pi\tau}  
-4(\mathcal{T}^{GG}_{12}){}^{\lambda}_{\mu\nu}{}^\rho_{\pi\tau}  
-2(\mathcal{T}^{GG}_{16}){}^{\lambda}_{\mu\nu}{}^\rho_{\pi\tau}  
         \\ & && \quad + 8(\mathcal{T}^{GG}_{18}){}^{\lambda}_{\mu\nu}{}^\rho_{\pi\tau}  
         + 4( \xi -1)(\mathcal{T}^{GG}_{21}){}^{\lambda}_{\mu\nu}{}^\rho_{\pi\tau}  
     + \frac{2}{{{D}}-2} (\mathcal{T}^{GG}_{22}){}^{\lambda}_{\mu\nu}{}^\rho_{\pi\tau}  \bigg].
\end{alignat}
\end{subequations}

The ghost propagator is given by:
\begin{equation}
    \label{fig:FRpghEH}
    \vcenter{\hbox{\includegraphics[scale=0.7]{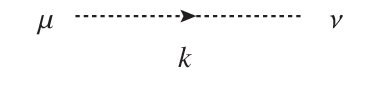}}}
    \frac{i}{k^{2}} \eta^{\mu \nu} 
\end{equation}

\subsubsection{Vertices}
The Feynman rule for the vertex $ (\mathcal{G} \mathcal{G} \bar{h}) $ (same of the usual vertex $ (GG h)$) is given by
\begin{equation}
    \label{fig:vertexbhGG}
    \vcenter{\hbox{\includegraphics[scale=0.7]{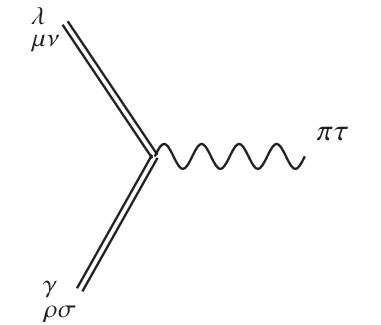}}}
i \kappa (\tensor{{N}}{_{\mu \nu}^{\lambda}_{\rho \sigma}^{\gamma}_{\pi \tau} }
    +
    \tensor{{N}}{_{\rho \sigma}^{\gamma}_{\mu \nu}^{\lambda}_{\pi \tau} }).
\end{equation}
The vertex $ ( \bar{G} \mathcal{G} \mathfrak{h} )$ has a similar structure: 
\begin{equation}
    \label{fig:vertexbGGh}
    \vcenter{\hbox{\includegraphics[scale=0.7]{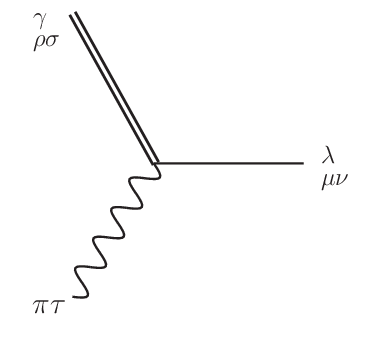}}}
    2i\tensor{{N}}{_{\mu \nu}^{\lambda}_{\rho \sigma}^{\gamma}_{\pi \tau} },
\end{equation}
where 
\begin{equation}\label{eq:vv3}
    \begin{split}
        \tensor{{N}}{_{\mu \nu}^{\lambda}_{\rho \sigma}^{\gamma}_{\pi \tau} } \equiv{}& \frac{1}{2} \frac{\delta \tensor{M}{ _{\mu \nu}^{\lambda}_{\rho \sigma}^{\gamma}}  (h)}{\delta h^{\pi \tau} }
        \\
        ={}&
\frac{1}{8 } \left \{ \left [ \left (
                \frac{\delta^{\gamma}_{\rho } \delta^{\pi}_{\nu } \delta^{\tau}_{\sigma } \delta^{\lambda}_{ \mu
            }}{{{D}}-1}-\delta^{\lambda}_{\rho } \delta^{\pi}_{\nu } \delta^{\tau }_{\sigma } \delta^{\gamma }_{\mu }
    + \pi \leftrightarrow \tau \right )  + \mu \leftrightarrow \nu \right ]  + \rho \leftrightarrow \sigma
\right \}  
    ,
\end{split}
\end{equation}
where $ \mu \leftrightarrow \nu $ denotes an index permutation.
Note that, these vertices are momenta independent.  

Now, we can consider the interactions that arises from the Faddeev-Popov action, which are momenta dependent. 
First, we have the gauge-fixing Lagrangian \eqref{eq:aa4}. The interaction $ \bar{\mathfrak{h}} \mathfrak{h} \mathfrak{h} $ comes from the partial derivatives:
\begin{equation}\label{eq:lowaa4}
    -\frac{\kappa }{2 \xi} \bar{\mathfrak{h}}_{\nu \beta} \partial_{\mu} \mathfrak{h}^{\mu \nu} \partial_{\alpha} \mathfrak{h}^{\alpha \beta}. 
\end{equation}
This leads to the following vertex (all momenta flows inwards):
\begin{equation}
    \label{fig:vertexbhhh}
     \vcenter{\hbox{\includegraphics[scale=0.7]{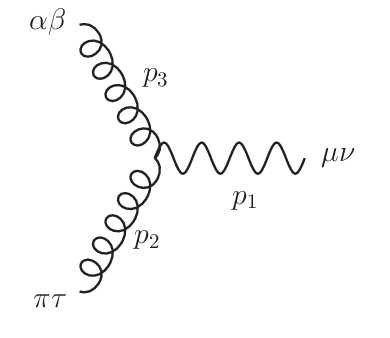}}}
     \begin{aligned}
         - \frac{i\kappa}{8 \xi }  & 
         \big[p_3^{\alpha}\left[\eta^{\nu\beta} \left(p_2^{\pi}\eta^{\mu\tau}+p_2^{\tau}\eta^{\mu\pi}\right)+\eta^{\mu\beta}\left(p_2^{\pi}\eta^{\nu\tau}+p_2^{\tau}\eta^{\pi\nu}\right)\right]\\
                                                 &  -p_3^{\beta}\left[p_2^{\pi} \left(\eta^{\mu\tau}\eta^{\alpha\nu}+\eta^{\mu\alpha}\eta^{\nu\tau}\right)+p_2^{\tau}\left(\eta^{\mu\alpha} \eta^{\pi\nu}+\eta^{\mu\pi}\eta^{\alpha\nu}\right)\right]\big].
\end{aligned}
    \end{equation}

    From the covariant derivatives (we have set $ {{D}} =4$)
\begin{equation}\label{eq:covDofmfrakh}
    \bar{\mathsf{D}}_{\mu} \mathfrak{h}^{\mu \nu} = 
  \partial_{\mu}\mathfrak{h}^{\mu\nu}
  + \bar{G}{}_{\mu\rho}{}^{\nu}\mathfrak{h}^{\mu\rho}
 - \frac{2}{3}\bar G_{\lambda \rho}{}^\lambda \mathfrak{h}^{\rho\nu},
\end{equation}
we get the interaction terms such as \( \bar{G} \mathfrak{h} \mathfrak{h} \). The last term in Eq.~\eqref{eq:covDofmfrakh} appears due to the tensor density nature of the Goldberg metric.
These terms yields the vertex
\begin{equation}
    \label{fig:vertexbGhh}
    \vcenter{\hbox{\includegraphics[scale=0.7]{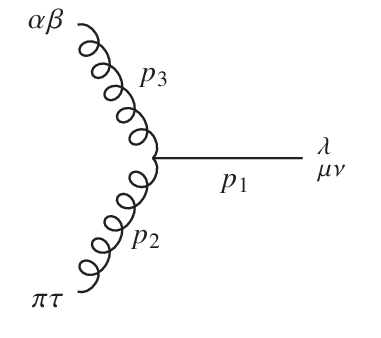}}}
    \begin{aligned}
    -    \frac{1}{4 \xi } & \big\{ 
(p_2^{\pi } \eta^{\lambda \tau }+\eta^{\pi \lambda }p_2^{\tau }) (\delta^\alpha_\nu \delta^\beta_\mu+\delta^\alpha_\mu \delta^\beta_\nu)
   \\ &  
   \quad +   (p_3^{\alpha } \eta^{\beta \lambda}+p_3^{\beta } \eta^{\alpha \lambda })(\delta^\pi_\nu\delta^\tau_\mu+\delta^\pi_\mu \delta^\tau_\nu)
                                                                                                                \\ & - \frac{1}{3}\big[ 
{p_2}^{\pi } (\eta^{\alpha \tau } \delta^\beta_\nu \delta_\mu^\lambda+\delta^\alpha_\nu \eta^{\beta \tau } \delta_\mu^\lambda+\eta^{\alpha \tau } \delta^\beta_\mu \delta^\lambda_\nu+\delta^\alpha_\mu \eta^{\beta \tau } \delta^\lambda_\nu)
                                                                                                                \\ & \qquad +{p_2}^{\tau } (\eta^{\pi \beta } \delta^\alpha_\nu \delta_\mu^\lambda+\eta^{\pi \alpha } \delta^\beta_\nu \delta_\mu^\lambda+\eta^{\pi \beta } \delta^\alpha_\mu \delta^\lambda_\nu+\eta^{\pi \alpha } \delta^\beta_\mu \delta^\lambda_\nu) \\ & \qquad +{p_3}^{\beta } (\delta^\pi_\nu \eta^{\alpha \tau } \delta_\mu^\lambda+\eta^{\pi \alpha } \delta_\mu^\lambda \delta^\tau_\nu+\delta^\pi_\mu \eta^{\alpha \tau } \delta^\lambda_\nu+\eta^{\pi \alpha } \delta^\lambda_\nu \delta^\tau_\mu) \\ & \qquad +{p_3}^{\alpha } (\delta^\pi_\nu \eta^{\beta \tau } \delta_\mu^\lambda+\eta^{\pi \beta } \delta_\mu^\lambda \delta^\tau_\nu+\delta^\pi_\mu \eta^{\beta \tau } \delta^\lambda_\nu+\eta^{\pi \beta } \delta^\lambda_\nu \delta^\tau_\mu)
\big]\big\}.
     \end{aligned}
\end{equation}

Finally, we have the interaction terms that come from the ghost Lagrangian \eqref{eq:aa5}. Considering only the partial-derivative terms, we obtain the following contributions:
\begin{equation}\label{eq:pdgh}
\kappa c^{\star \mu} \left[
    \eta_{\mu \nu}
    \left(\bar{\mathfrak{h}}^{\alpha \beta}   \partial_{\alpha} \partial_{\beta} 
      + 
    \frac{1}{2} \bar{\mathfrak{h}}_{\rho}^{\rho}   \partial_{\alpha} \partial^{\alpha }  \right)
    +  
     \bar{\mathfrak{h}}_{\mu \nu} \partial_{\alpha} \partial^{\alpha} 
 \right]c^{\nu}
+ 
\kappa \eta_{\beta \nu} c^{\star \beta} \left[ (\partial_{\mu} \bar{\mathfrak{h}}^{\mu \lambda} )\partial_{\lambda} c^{\nu}  -    (\partial_{\mu} \partial_{\lambda} \bar{\mathfrak{h}}^{\mu \nu}) c^{\lambda}
-    (\partial_{\mu}  \bar{\mathfrak{h}}^{\mu \nu}) \partial_{\lambda}c^{\lambda}\right]
    ,
\end{equation}
where we used that $ \sqrt{\bar{h}} = 1 + \kappa \bar{\mathfrak{h}}_{\rho}^{\rho} /2 + O ( \kappa ^2 )$ .
From the terms that arises from the covariant derivative
\begin{equation}\label{eq:cvgh}
    \begin{split}
        \mathsf{D}_{\lambda} \left(\mathsf{D}_{\gamma} d^{\nu}\right) ={}& \partial_{\lambda} \left\{ \left[G_{\gamma\alpha}{}^\nu - \frac{1}{3} (\delta^\nu_\gamma G_{\sigma\alpha}{}^\sigma + \delta^\nu_\alpha G_{\sigma\gamma}{}^\sigma) \right]d^{\alpha}  \right\} \\ & +  \left[G_{\lambda\alpha}{}^\nu - \frac{1}{3} \left(\delta^\nu_\lambda G_{\sigma\alpha}{}^\sigma + \delta^\nu_\alpha G_{\sigma\lambda}{}^\sigma\right)  \right]\partial_{\gamma}d^{\alpha} 
                                                      \\ & - \left[G_{\lambda\gamma}{}^\alpha - \frac{1}{3} (\delta^\alpha_\lambda G_{\sigma\gamma}{}^\sigma + \delta^\alpha_\gamma G_{\sigma\lambda}{}^\sigma)\right] 
    \partial_{\alpha} d^{\nu}
    \end{split}
\end{equation}
and the terms of the ghost Lagrangian~\eqref{eq:aa5}, $ \bar{R}_{\nu \mu } ( \bar{G} ) = - \tensor{G}{_{\mu \nu}^{\lambda}_{, \lambda}} -(\tensor{G}{_{\nu \rho}^{\rho}_{, \mu}} - \tensor{G}{_{\mu \rho}^{\rho}_{, \nu}} )/3+ O( \bar{G}^{2} )$ and $ \mathsf{X}_{\mu \nu} $, we get all the contributions to the vertex $ \bar{G} \mathfrak{h} \mathfrak{h} $. 

Above interactions terms lead to the ghost vertices:
\begin{equation}
    \label{fig:vertexbpbcc}
     \vcenter{\hbox{\includegraphics[scale=0.7]{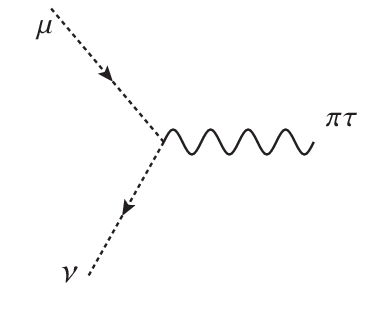}}}
     \begin{aligned}
         \frac{i \kappa }{2}  
         \big[ & 
             p_1^{\tau} (p_3^{\pi} \eta^{\nu\mu}-p_1^{\mu} \eta^{\pi\nu}-p_3^{\mu} \eta^{\pi\nu})  -p_1^{\pi} (p_1^{\mu} \eta^{\nu\tau}+p_3^{\mu} \eta^{\nu\tau}-p_3^{\tau} \eta^{\nu\mu}) \\ & +2 p_3^{\pi} p_3^{\tau} \eta^{\nu\mu}+(p_3)^{2} (\eta^{\pi\tau} \eta^{\nu\mu}-\eta^{\pi\mu} \eta^{\nu\tau}-\eta^{\pi\nu} \eta^{\mu\tau})
\big]
\end{aligned}
    \end{equation}
and
\begin{equation}
    \label{fig:vertexbGbcc}
     \vcenter{\hbox{\includegraphics[scale=0.7]{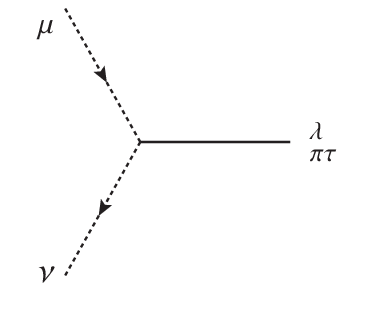}}}
     \begin{aligned}
         \frac{1}{3}  
         \big[ & 
             \delta_{\tau }^{\lambda} (p_3{}_{\pi } \eta_{\mu \nu }-\eta_{\pi \nu } p_3{}_{\mu }+\eta_{\pi \mu } p_3{}_{\nu }) 
            \\ & -3 \delta_{\nu }^{\lambda} (p_3{}_{\pi } \eta_{\mu \tau }-\eta_{\pi \tau }p_3{}_{\mu }+\eta_{\pi \mu } p_3{}_{\tau }) \\ & -\delta_{\pi}^{\lambda } (p_3{}_{\mu } \eta_{\nu \tau }-p_3{}_{\nu } \eta_{\mu \tau }-p_3{}_{\tau } \eta_{\mu \nu })
      \big].
\end{aligned}
\end{equation}

Quartic vertices are omitted, since they give rise only to tadpole-like contributions, which vanish in dimensional regularization.

\subsection{Self-Energies}

Using the Feynman rules obtained above, we can compute the one-loop diagrams in Fig.~\ref{fig:1}(a) using the Passarino-Veltman approach. The divergent part of the self-energies $ \bar{\mathfrak{h}} \bar{\mathfrak{h}}$, $ \bar{G} \bar{\mathfrak{h}}$, $\bar{\mathfrak{h}} \bar{G}$ and $\bar{G} \bar{G}$ reads

\begin{subequations}\label{eq:SEtable}
\begin{align}\label{eq:SEhhresult}
    \Pi_{\mu \nu \alpha \beta}(k) ={}& \frac{\kappa^{2} k^{4}}{16 \pi^{2} \epsilon} \sum_{J=I}^{V} \sum_{i=1}^{5} a_{i}^{(J)}  ( \xi) \mathcal{T}^{(i)}{}_{\mu \nu \alpha \beta} (k),\\ \label{eq:SEGhresult}
    \tensor{\Pi}{_{\mu \nu}^{\lambda}_{ \pi \tau }}(k) ={}& \frac{i\kappa k^{2}}{16 \pi^{2} \epsilon} \sum_{J=I}^{V} \sum_{i=1}^{12} b_{i}^{(J)}  ( \xi) (\mathcal{T}^{Gh}_i)\tensor{}{_{\mu \nu}^{\lambda}_{ \pi \tau }}(k)
    ,\\
    \tensor{\Pi}{_{\mu \nu}_{\pi \tau }^{\rho }} (k)
    ={}& - 
    \tensor{\Pi}{_{ \pi \tau }^{\rho }_{\mu \nu}} (k)
\\ \label{eq:SEGGresult}
    \tensor{\Pi}{_{\mu \nu}^{\lambda}_{ \pi \tau }^{\rho}} (k) 
    ={}&
    \frac{ k^{2}}{16 \pi^{2} \epsilon} \sum_{J=I}^{V} \sum_{i=1}^{22} c_{i}^{(J)}  ( \xi) (\mathcal{T}^{GG}_i)\tensor{}{_{\mu \nu}^{\lambda}_{\pi \tau }^{\rho} }(k);
\end{align}
\end{subequations}
where the symbolic summation over $J$ represents the five diagrams ($I, II, III, IV, V$). The coefficients $ a_{i}^{(J)} ( \xi )$, $ b_{i}^{(J)} ( \xi )$ and $ c_{i}^{(J)} ( \xi )$ are shown in the Tables~\ref{table:hh},~\ref{table:Gh} and \ref{table:GG}.
\begin{table*}[ht]
    \centering
    \caption{The coefficients $  a_{m}^{(J)} $ (see Eq.~\eqref{eq:SEhhresult}) for the divergent part of the diagrams in Fig.~\ref{fig:1}(a) decomposed in the basis \eqref{eq:32}.}
\label{table:hh}
\begin{tblr}{c|ccccc}
    \toprule
    $m $ & $I$ & $II$& $III$ & $IV$& $V$ \\ \midrule
 $1   $&$0 $&$ 0 $&$ 0 $&$  \dfrac{11 \xi ^2-2 \xi +1}{60}  $&$ -\dfrac{7}{15}  $\\
 $3   $&$0 $&$ 0 $&$ 0 $&$ - \dfrac{\xi ^2+3 \xi -4}{120} $&$ -\dfrac{1}{60}  $\\
 $3   $&$0 $&$ 0 $&$ 0 $&$  \dfrac{-2 \xi ^2+14 \xi +23}{240}  $&$ -\dfrac{1}{60}   $\\
 $4   $&$0 $&$ -\dfrac{\xi }{24}  $&$ -\dfrac{\xi }{24}$&$   - \dfrac{13 \xi ^2-26 \xi +43}{240}  $&$- \dfrac{13}{120}   $\\
 $5   $&$0 $&$ \dfrac{\xi}{48}  $&$ \dfrac{\xi }{48}  $&$ \dfrac{12 \xi ^2-54 \xi +37}{240}  $&$ -\dfrac{1}{240}  $\\
\bottomrule
\end{tblr}
\end{table*}

\begin{table*}[ht]
    \centering
    \caption{The coefficients $  b_{m}^{{(J)}} $ (see Eq.~\eqref{eq:SEGhresult}) for the divergent part of the diagrams in Fig.~\ref{fig:1}(b) decomposed in the basis \eqref{eq:TensHp}.}
\label{table:Gh}
\begin{tblr}{c|ccccc}
    \toprule
    $m $ & $I$ & $II$& $III$ & $IV$& $V$ \\ \midrule
    $1 $&$0 $&$ - \dfrac{\xi +3}{18} $&$ -\dfrac{ \xi }{12} $&$ - \dfrac{8 \xi ^2-31 \xi +23}{120} $&$ \dfrac{1}{6} $\\
    $2 $&$0 $&$  \dfrac{\xi -2}{12} $&$ -\dfrac{5 \xi }{12} $&$ - \dfrac{14 \xi ^2+37 \xi +79}{120} $&$ 0 $\\
    $3 $&$0 $&$ - \dfrac{7 \xi -5}{48} $&$ -\dfrac{ \xi }{24} $&$  \dfrac{4 \xi ^2-8 \xi +9}{60} $&$ \dfrac{1}{24} $\\
    $4 $&$0 $&$  \dfrac{3 \xi -1}{24} $&$ 0 $&$ - \dfrac{\xi ^2-2 \xi +6}{30} $&$ -\dfrac{1}{12} $\\
    $5 $&$0 $&$ - \dfrac{\xi -1}{12} $&$ \dfrac{ \xi }{12} $&$  \dfrac{2 \xi ^2-29 \xi +107}{120} $&$ 0 $\\
    $6 $&$0 $&$  \dfrac{\xi +1}{24} $&$ 0 $&$  \dfrac{\xi ^2+8 \xi -9}{120} $&$ \dfrac{1}{36} $\\
    $7 $&$0 $&$  \dfrac{5 \xi -3}{36} $&$ 0 $&$ - \dfrac{14 \xi ^2-3 \xi +9}{60} $&$ \dfrac{1}{18} $\\
    $8 $&$0 $&$ - \dfrac{\xi -1}{6} $&$ \dfrac{ \xi }{6} $&$  \dfrac{2 \xi ^2+31 \xi -43}{60} $&$ - \dfrac{2}{3}  $\\
    $9 $&$0 $&$ - \dfrac{\xi -1}{6} $&$ 0 $&$  \dfrac{11 \xi ^2-17 \xi +6}{60} $&$ 0 $\\
    $10 $&$0 $&$ 0 $&$ 0 $&$ 0 $&$ 0 $\\
    $11 $&$0 $&$ - \dfrac{\xi -3}{24} $&$ -\dfrac{ \xi }{12} $&$  \dfrac{41 \xi ^2-27 \xi +46}{120} $&$ \dfrac{5 }{12} $\\
    $12 $&$0 $&$  \dfrac{\xi -1}{6} $&$ 0 $&$ - \dfrac{9 \xi ^2-13 \xi +4}{120} $&$ 0 $\\
\bottomrule
\end{tblr}
\end{table*}

\begin{table*}[ht]
    \centering
    \caption{The coefficients $  c_{m}^{{(J)}} $ (see Eq.~\eqref{eq:SEGGresult}) for the divergent part of the diagrams in Fig.~\ref{fig:1}(c) decomposed in the basis \eqref{eq:TensGG}.}
\label{table:GG}
\begin{tblr}{c|ccccc}
    \toprule
    $m $ & $I$ & $II$& $III$ & $IV$& $V$ \\ \midrule
    $1 $&$ -\dfrac{2}{9  } $&$ \dfrac{5 \xi +9}{18  } $&$ \dfrac{10 \xi ^2-9 \xi -9}{24  } $&$ -\dfrac{26 \xi ^2-32 \xi +1}{60  } $&$ \dfrac{2}{9  } $\\
    $2 $&$ 0 $&$ -\dfrac{\xi -1}{6  } $&$ -\dfrac{18 \xi ^2+7 \xi -9}{24  } $&$ \dfrac{42 \xi ^2-24 \xi +7}{60  } $&$ 0 $\\
    $3 $&$ \dfrac{1}{12  } $&$ \dfrac{\xi -1}{12  } $&$ -\dfrac{6 \xi ^2-19 \xi +13}{24  } $&$ \dfrac{8 \xi ^2-16 \xi +13}{30  } $&$ -\dfrac{1}{12  } $\\
    $4 $&$ \dfrac{1}{3  } $&$ \dfrac{\xi -11}{6  } $&$ \dfrac{10 \xi ^2+3 \xi +11}{24  } $&$ -\dfrac{26 \xi ^2+8 \xi -9}{60  } $&$ -\dfrac{1}{3  } $\\
    $5 $&$ \dfrac{1}{9  } $&$ \dfrac{7-4 \xi }{9  } $&$ \dfrac{-4 \xi ^2+\xi +3}{12  } $&$ \dfrac{12 \xi ^2-9 \xi -8}{30  } $&$ -\dfrac{1}{9  } $\\
    $6 $&$ 0 $&$ 0 $&$ 0 $&$ 0 $&$ 0 $\\
    $7 $&$ 0 $&$ 0 $&$ 0 $&$ 0 $&$ 0 $\\
    $8 $&$ 0 $&$ 0 $&$ 0 $&$ 0 $&$ 0 $\\
    $9 $&$ 0 $&$ 0 $&$ 0 $&$ 0 $&$ 0 $\\
    $10 $&$ 0 $&$ -\dfrac{2 \xi }{9  } $&$ -\dfrac{\xi ^2-3 \xi +2}{3  } $&$ \dfrac{14 \xi ^2-28 \xi +19}{30  } $&$ 0$\\
    $11 $&$ 0 $&$ -\dfrac{\xi -1}{3  } $&$ \dfrac{10 \xi ^2-13 \xi +19}{12  } $&$ -2\dfrac{ 4 \xi ^2-13 \xi +9}{15  } $&$ 0 $\\
    $12 $&$ \dfrac{2}{3  } $&$ \dfrac{2}{3  } $&$ -\dfrac{(\xi -1) \xi }{3  } $&$ \dfrac{14 \xi ^2-18 \xi +29}{30  } $&$- \dfrac{2}{3  } $\\
    $13 $&$ -\dfrac{4}{9  } $&$ \dfrac{3 \xi -7}{9  } $&$ \dfrac{4 \xi ^2-7 \xi +3}{6  } $&$ -\dfrac{13 (\xi -1)^2}{15  } $&$ \dfrac{4}{9  } $\\
    $14 $&$ 0 $&$ \dfrac{\xi -1}{6  } $&$ -\dfrac{\xi ^2-2 \xi +3}{3  } $&$ \dfrac{7 \xi ^2-34 \xi +27}{30  } $&$ 0 $\\
    $15 $&$ \dfrac{4}{9  } $&$ \dfrac{8 (\xi -2)}{9  } $&$ \dfrac{-6 \xi ^2+11 \xi +3}{6  } $&$ \dfrac{14 (\xi -1)^2}{15  } $&$ 
    -\dfrac{4}{9}$\\
    $16 $&$ 0 $&$ \dfrac{5 \xi +1}{3  } $&$ \dfrac{18 \xi ^2-9 \xi +7}{12  } $&$ \dfrac{2 \xi ^2-4 \xi +27}{30  } $&$ 0 $\\
    $17 $&$ -\dfrac{2}{3  } $&$ -(\xi -3) $&$ \dfrac{2 \xi ^2-3 \xi +5}{6  } $&$ -\dfrac{3 \xi ^2-11 \xi +8}{15  } $&$ \dfrac{2}{3}$\\
    $18 $&$ 0 $&$ -\dfrac{2 (\xi +5)}{3  } $&$ \dfrac{-2 \xi ^2+\xi -7}{2  } $&$ \dfrac{4 \xi ^2-13 \xi -16}{15  } $&$ 0 $\\
    $19 $&$ 0 $&$ \dfrac{16}{9  } $&$ \dfrac{\xi ^2-4 \xi +1}{3  } $&$ -\dfrac{8 \xi ^2-11 \xi +8}{15  } $&$ 0 $\\
    $20 $&$ 0 $&$ 0 $&$ 0 $&$ 0 $&$ 0 $\\
    $21 $&$ 0 $&$ 0 $&$ 0 $&$ 0 $&$ 0 $\\
    $22 $&$ \dfrac{1}{6  } $&$ -\dfrac{\xi +1}{2  } $&$ \dfrac{\xi ^2-\xi +2}{3  } $&$ \dfrac{3 \xi ^2-\xi +3}{15  } $&$ -\dfrac{1}{6  } $\\
\bottomrule
\end{tblr}
\end{table*}

The coefficients $b_{10} $, $c_{6} $--$c_{9}$, $ c_{20} $ and $c_{21} $ vanish, since non-local terms, as 
\begin{equation}\label{eq:appV26}
    k^2 \frac{k_{\mu} k_{\nu} k_{\alpha} k_{\beta}}{ k^4} \xrightarrow[\text{space} ]{\text{position}} \frac{\partial_{\mu} \partial_{\nu} \partial_{\alpha } \partial_{\beta}}{\partial^{2}} \quad \text{and} \quad 
    k^2 \frac{k_{\mu} k_{\nu} k_{\alpha} k_{\beta} k^{\lambda} k^{\rho} \partial^{\lambda} \partial^{\rho} }{ k^6} \xrightarrow[\text{space} ]{\text{position}} \frac{\partial_{\mu} \partial_{\nu} \partial_{\alpha } \partial_{\beta}}{\partial^{2} \partial^{2} } ,
\end{equation}
would arise. Such terms are forbidden due to the locality of gauge transformation and the corresponding BRST transformations.

\subsection{Proper self-energy}

Setting the independent field $ \bar{G} $ on-shell in Eq.~\eqref{eq:31}, we obtain the corresponding proper effective Lagrangian 
\begin{equation}\label{eq:31app}
    \mathcal{L}^{\text{I}}_{\text{eff}} \left( \bar{\mathfrak{h}} , \mathcal{G} (\bar{\mathfrak{h} })  \right) = 
    \frac{1}{2} \left ( \bar{\mathfrak{h}}^{\mu \nu} \Pi_{\mu \nu \alpha \beta} \bar{\mathfrak{h}}^{\alpha \beta} + 2\tensor{\mathcal{G}}{_{\alpha \beta}^{\sigma}} \tensor{\Pi}{^{\alpha \beta}_{\sigma }_{\mu \nu} } \bar{\mathfrak{h}}^{\mu \nu} + \tensor{\mathcal{G}}{_{\mu \nu}^{\rho}} \tensor{\Pi}{^{\mu \nu}_{\rho}^{\alpha \beta}_{\sigma}} \tensor{\mathcal{G}}{_{\alpha \beta}^{\sigma}}\right )
    =
    \frac{1}{2} \bar{\mathfrak{h}}^{\mu \nu} \Pi_{\mu \nu \alpha \beta}^{\text{I}}  \bar{\mathfrak{h}}^{\alpha \beta} 
    .
\end{equation}
Taking derivatives with respect the field $ \bar{\mathfrak{h}} $, we obtain the two-point function:
\begin{equation} \label{eq:defpSE}
    {\Pi}_{\mu \nu \alpha \beta}^{\text{I}} (k) 
=
{\Pi}_{\mu \nu \alpha \beta} (k) 
+ 
\frac{\delta \mathcal{G}^{\pi \tau}_{\lambda}} { \delta \bar{\mathfrak{h}}^{\mu \nu} }
{\Pi}_{\pi \tau }{}^{\lambda}{}_{ \alpha \beta}(k)
+ 
    {\Pi}^{\mu \nu \rho \sigma}{}_{\gamma}  (k) 
\frac{\delta \mathcal{G}_{\rho \sigma}^{\gamma}} { \delta \bar{\mathfrak{h}}^{\alpha \beta } }
+  
\frac{\delta \mathcal{G}^{\pi \tau}_{\lambda}} { \delta \bar{\mathfrak{h}}^{\mu \nu} }
{\Pi}_{\pi \tau }{}^{\lambda}{}_{ \rho \sigma}{}^{\gamma} (k) 
\frac{\delta \mathcal{G}^{\rho \beta '}_{\gamma}} { \delta \bar{\mathfrak{h}}^{\alpha \beta } },
\end{equation}
where the derivatives are taken with $ \bar{\mathfrak{ h}} =0$. From the equation of motion~\eqref{eq:29}, we find that the proper self-energy can be written as 
\begin{equation} \label{eq:defpSEcomputed}
    {\boldsymbol{\Pi}}^{\text{I}} 
=
{\boldsymbol{\Pi}}_{\bar{\mathfrak{h}} \bar{\mathfrak{h}}} 
+ 
\kappa \boldsymbol{B} \boldsymbol{M}^{-1} (\eta)
{\boldsymbol{\Pi}}_{\bar{G} \bar{\mathfrak{h}}} 
- \kappa  
{\boldsymbol{\Pi}}_{\bar{\mathfrak{h}} \bar{G}}   
    \boldsymbol{M}^{-1}(\eta ) \boldsymbol{C} 
    -  \kappa^{2}  
\boldsymbol{ B} \boldsymbol{M}^{-1} (\eta )
{\boldsymbol{\Pi}}_{\bar{G} \bar{G}} 
\boldsymbol{M}^{-1}(\eta ) \boldsymbol{C},
\end{equation}
where we used the matrices $ 
 {\boldsymbol{\Pi}}_{\bar{\mathfrak{h}} \bar{\mathfrak{h}}}$,  
$
{\boldsymbol{\Pi}}_{\bar{G} \bar{\mathfrak{h}}}$,  
$
{\boldsymbol{\Pi}}_{\bar{\mathfrak{h}} \bar{G}}$ and $
{\boldsymbol{\Pi}}_{\bar{G} \bar{G}} 
$ that denote respectively the self-energies $ \bar{\mathfrak{h}} \bar{\mathfrak{h}}$, $ \bar{G} \bar{\mathfrak{h}}$, $\bar{\mathfrak{h}} \bar{G}$ and $\bar{G} \bar{G}$ and $ \boldsymbol{B} $, $ \boldsymbol{C} $ and $ \boldsymbol{M} (\eta)= \boldsymbol{D}  $ defined in Eq.~\eqref{eq:matrix}.

Using Eq.~\eqref{eq:matrix}, we obtain that 
\begin{equation}\label{eq:explicitONSHELL}
    \begin{split}
    [ \boldsymbol{B} \boldsymbol{M}^{-1}(\eta )]{}^{\pi \tau\ \mu \nu}{}_{\lambda } 
    ={}&
    - [ \boldsymbol{M}^{-1} ( \eta ) \boldsymbol{C} ]{}^{ \mu \nu}{}_{ \lambda }{}^{\pi \tau} \\
    ={}&\frac{i}{4} \left[k^{\pi } (\delta_\lambda^\nu \eta^{\mu \tau }+
   \delta_\lambda^\mu \eta^{\nu \tau })+k^{\tau }(\eta^{\pi \nu }
    \delta_\lambda^\mu+\eta^{\pi \mu }  \delta_\lambda^\nu)
-k_{\lambda} \left(\eta^{\pi \nu }  \eta^{\mu \tau }
+\eta^{\pi \mu }  \eta^{\nu \tau }
    +\frac{2 \eta^{\pi
   \tau }  \eta^{\mu \nu }}{2-{{D}}}\right)\right].
\end{split}
\end{equation}
Substituting this and the self-energies computed above in  Eq.~\eqref{eq:defpSEcomputed}, we obtain the divergent part of the proper self-energy: 
 \begin{equation}\label{eq:resultforpSE}
     \begin{split}
         \Pi^{\text{I div}}_{\mu \nu \alpha \beta} ={}&
         \frac{\kappa^{2} k^{4}}{16 \pi^{2} \epsilon} \sum_{J=I}^{V} \sum_{m=1}^{5} \mathcal{C}_{m}^{(J)} ( \xi ) \mathcal{T}^{(m)}{}_{\mu \nu \alpha \beta} (k)
         =
     \frac{\kappa^{2} k^{4}}{16 \pi^{2} \epsilon} \sum_{m=1}^{5} C_{m} ( \xi ) \mathcal{T}^{(m)}{}_{\mu \nu \alpha \beta} (k),
     \end{split}
 \end{equation} 
 where 
the coefficients $ \mathcal{C}_{m}^{(J)} $ are presented in Table~\ref{table:PSE}. Summing all entries of the $m$-row yields the coefficient ${C}_{m} ( \xi )$ shown in Eq.~\eqref{eq:35}. 
 \begin{table*}[ht]
    \centering
    \caption{The coefficients $\mathcal{C}_{m}^{{(J)}} $ (see Eq.~\eqref{eq:resultforpSE}) for the divergent part of the proper self-energy  decomposed in the basis \eqref{eq:32}. The total of each row yields $ {C}_{m} $ of Eq.~\eqref{eq:35}.}
\label{table:PSE}
\begin{tblr}{c|rcccc}
    \toprule
    $m $ & $I$ & $II$ & $III$& $IV$& $V$ \\ \midrule
    $1$ & $\dfrac{1}{6} $&$ \dfrac{7}{6} $&$  \dfrac{10 \xi ^2-7 \xi +13}{12} $&$  \dfrac{2 \xi ^2-9\xi +13}{12} $&$ -\dfrac{4}{5} $\\
$2$ & $\dfrac{1}{24} $&$ - \dfrac{3 \xi +2}{12} $&$  \dfrac{-\xi ^2+23 \xi -28}{48} $&$  \dfrac{\xi ^2-15 \xi +21}{48} $&$ -\dfrac{17}{120} $\\
$3$ &$ \dfrac{1}{24} $&$ \dfrac{\xi +3}{4} $&$  \dfrac{\xi ^2-15 \xi +19}{24} $&$  \dfrac{10 \xi^2+6 \xi -21}{48} $&$ -\dfrac{7}{120} $\\
$4$ & $\dfrac{1}{12} $&$ \dfrac{23-\xi }{24} $&$  \dfrac{8 \xi ^2-17 \xi +31}{48} $&$  \dfrac{16\xi ^2-13 \xi +3}{48} $&$ -\dfrac{2}{5} $\\
$5$ & $-\dfrac{1}{48} $&$ - \dfrac{11 \xi +38}{48} $&$  \dfrac{-8 \xi ^2+30 \xi -33}{48} $&$  \dfrac{-4 \xi ^2-7 \xi +17}{48} $&$ \dfrac{7}{120} $\\
\bottomrule
\end{tblr}
\end{table*}

The Ward identity Eq.~\eqref{eq:34a} satisfied by the proper self-energy \eqref{eq:resultforpSE} implies that the coefficients $C_{i} ( \xi )$ are not independent. Indeed, imposing it to a general tensor \[ (k^{\mu} \eta^{\rho \nu} + k^{\nu} \eta^{\rho \mu} - k^{\rho} \eta^{\mu \nu} )\sum_{i=1}^{5} C_{i} \mathcal{T}^{(i)}_{\mu \nu \alpha \beta} (k)=0,\] one finds that 
\begin{equation}\label{eq:transversality}
    C_{1} = 4 ( C_{2} + C_{3} ), \quad C_{4 } = \frac{C_{1} }{2} \quad \text{and} \quad C_{5} = - C_{3},
\end{equation}
which are the same conditions satisfied by $ C_{i} ( \xi )$ in Eq.~\eqref{eq:35}.

\subsubsection{Counterterm Lagrangian}

The counterterm Lagrangian can be obtained from the divergent part of the proper effective Lagrangian~\eqref{eq:31app}. To connect these expressions, we note that the invariants constructed from
\begin{subequations}\label{eq:defRinhapp}
    \begin{align}\label{eq:36app}
        \bar{R} ={}& \kappa k^{2} L_{\alpha \beta} 
        \left ( \frac{1}{2} \eta^{\alpha \beta } \eta_{\pi \tau}  - I^{\alpha \beta}_{\pi \tau}  \right ) \bar{\mathfrak{ h}}^{\pi \tau} 
         + O ( \kappa^{2} ), \\
        \bar{R}_{\mu \nu} ={}& \frac{\kappa}{2} \left [ k_{\mu} k_{\nu} L_{\rho \sigma} - \frac{k^{2}}{2} \left ( L_{\mu \sigma } L_{\rho \nu} + L_{\mu \rho} L_{\sigma \nu}\right )\right ] 
        \left ( \frac{1}{2} \eta^{\rho \sigma } \eta_{\pi \tau}  - I^{\rho \sigma}_{\pi \tau}  \right ) \bar{\mathfrak{ h}}^{\pi \tau} 
        + O ( \kappa^{2} ); 
    \end{align}
\end{subequations}
where $ I_{\mu \nu}^{\alpha \beta} = ( \delta_{\mu}^{\alpha} \delta_{\nu}^{\beta} + \delta_{\mu}^{\beta} \delta_{\nu}^{\alpha } )/2$, are decomposed in the basis $\{ \mathcal{T}^{(i)} \}$ as 
\begin{subequations}\label{eq:invariants}
    \begin{align}
        \sqrt{- \bar{g} } \bar{R}^{2} ={}& \frac{1}{4} \kappa^{2} k^{4} \left( 4\mathcal{T}^{(1)} + \mathcal{T}^{(2)} + 2 \mathcal{T}^{(4)} \right)_{\mu \nu \alpha \beta} \bar{\mathfrak{h}}^{\mu \nu} \bar{\mathfrak{h}}^{\alpha \beta}, \\
        \sqrt{- \bar{g} } \bar{R}_{\mu \nu} \bar{R}^{\mu \nu}  ={}& \frac{1}{8} \kappa^{2} k^{4} \left ( 4 \mathcal{T}^{(1)} + \mathcal{T}^{(3)} + 2 \mathcal{T}^{(4)} - \mathcal{T}^{(5)} \right )_{\mu \nu \alpha \beta} \bar{\mathfrak{h}}^{\mu \nu} \bar{\mathfrak{h}}^{\alpha \beta} . 
\end{align}
\end{subequations}

The proper effective Lagrangian in Eq.~\eqref{eq:31app} can thus be written as
\begin{equation}\label{eq:idCRCRmunu}
    \frac{1}{2}   \frac{\kappa^{2} k^{4}}{16 \pi \epsilon} \left\{C_{2} ( \xi ) \left [ 4 \mathcal{T}^{(1)} + \mathcal{T}^{(2)} + 2 \mathcal{T}^{(4)}\right ]_{\mu \nu \alpha \beta} (k) 
    + 
    C_{3} ( \xi ) \left [ 4 \mathcal{T}^{(1)} + \mathcal{T}^{(3)} + 2 \mathcal{T}^{(4)} - \mathcal{T}^{(5)} \right ]_{\mu \nu \alpha \beta} (k)
\right\}
\bar{\mathfrak{h}}^{\mu \nu} \bar{\mathfrak{h}}^{\alpha \beta}
\end{equation}
and by using Eq.~\eqref{eq:invariants}, we obtain that
\begin{equation}\label{eq:38app}
    \left .\mathcal{L}_{\text{CT}}^{\text{I}} \right |_{\bar{G} = \mathcal{G}} = 
        \frac{\sqrt{- \bar{g}} }{16 \pi^{2} \epsilon} \left [   
            2C_{2} ( \xi ) \bar{R}^{2} + 
            4C_{3} ( \xi )
\bar{R}_{\mu \nu} \bar{R}^{\mu \nu}\right ],
\end{equation}
which is equal to the result shown in Eq.~\eqref{eq:38}.

\section{Counterterms in the Yang-Mills theory} \label{sec:YM}
The analogous of the Eq.~\eqref{eq:210} in the Yang-Mills theory is given by:
\begin{equation}\label{eq:old210}
    \int \mathop{\mathcal{D} \mathfrak{F}^{a \, \mu \nu}} \exp i \int \mathop{d^{} x} \left(\mathcal{L}^{\text{I}}_{\text{YM}} - \mathcal{L}^{\text{II}}_{{\text{YM}}} \right)= \exp \frac{i}{2}  \int \mathop{d^{4}x}  \left [ \bar{F}_{\mu \nu}^{a} - f_{\mu \nu}^{a} ( \bar{A} )\right ] \left [ \frac{1}{2} \left ( \bar{F}^{a \, \mu \nu} - f^{a \, \mu \nu} ( \bar{A} )\right ) - g f^{abc} \mathfrak{A}^{b \, \mu} \mathfrak{A}^{c \, \nu} \right ],
\end{equation}
where $ \bar{A} $, $ \bar{F} $ are respectively the background gauge and independent fields, $ \mathfrak{A} $, $ \mathfrak{F} $ are the corresponding quantum fields and $f_{\mu \nu}^{a} ( \bar{A} ) \equiv  \partial_{\mu} \bar{A}^{a}_{\nu} - \partial_{\nu} \bar{A}^{a}_{\mu} + g f^{abc} \bar{A}^{b}_{\mu } \bar{A}^{c}_{\nu}$ (for further details on the first order formulation of the Yang-Mills in the background field method, see Ref.~\cite{Brandt:2018wxe}).

Eq.~\eqref{eq:old210} shows that when the background field $ \bar{F}_{\mu \nu}^{a} $ is on-shell: $ \bar{F}_{\mu \nu}^{a} = f_{\mu \nu}^{a} ( \bar{A} ) $, the first and the second order Yang-Mills Lagrangians, in the background field method, turn out to be equivalent. Following the same reasoning in Sec. II, one can show that this also holds for the effective actions when the background field $ \bar{F}^{a}_{\mu \nu} $ is on-shell (and the source of the complementary field set to zero). 

\begin{figure}[ht]
    \includegraphics[width=0.618\textwidth]{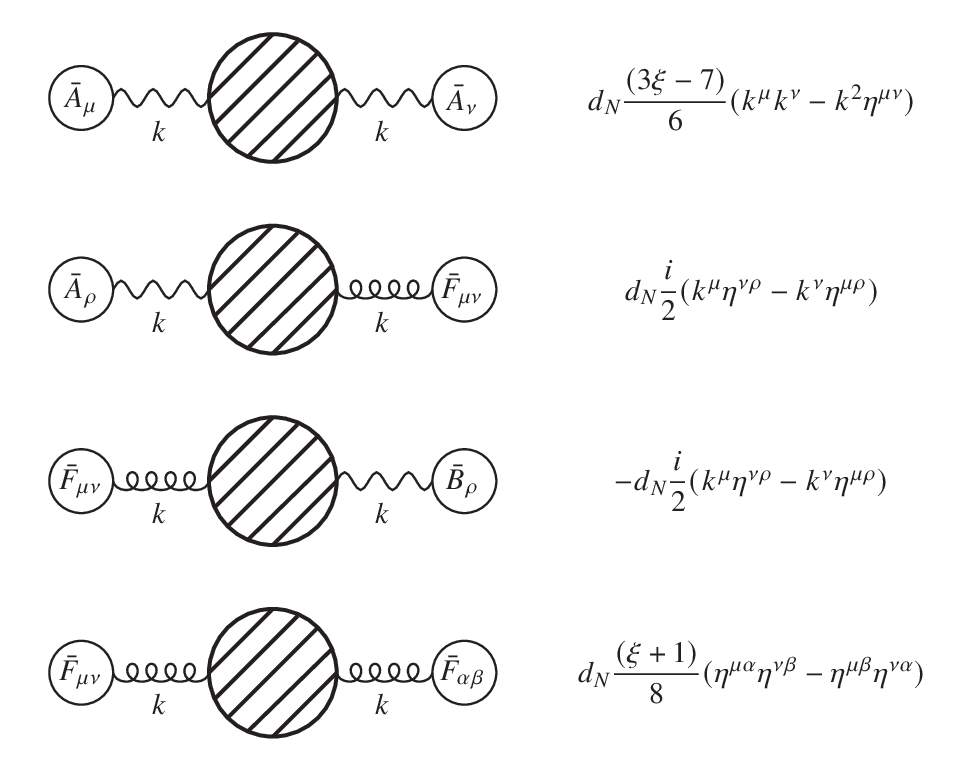}
    \caption{Divergent part of the background self-energies in the first-order Yang-Mills theory. The factor $ d_{N} $ is equal to $ iN g^{2} \delta^{ab} /16 \pi^{2} \epsilon$.}\label{fig:BSEFOYM}
\end{figure}

Now, we consider the contributions to the gluon self-energy in the first order formulation \cite{Brandt:2018wxe} illustrated in Fig.~\ref{fig:BSEFOYM}. 
The divergent parts of these contributions yield the  first order counterterm Lagrangian 
\begin{equation}\label{eq:old224}
\mathcal{L}^{\text{I}}_{\text{CT}} = - \frac{1}{2} d_{N} \left [
    C_{\bar{F} \bar{F} } 
\bar{F}_{\mu \nu}^{a} \bar{F}^{a \, \mu \nu} 
+     2C_{ \bar{F}  \bar{f}} 
  \bar{F}_{\mu \nu}^{a} f^{a \, \mu \nu} ( \bar{A} )  + 
    C_{\bar{f} \bar{f}}
f_{\mu \nu}^{a} ( \bar{A} ) f^{a \, \mu \nu} (\bar{A} ) \right ],
\end{equation}
where 
$ d_{N} = g^{2} N /16 \pi^{2} \epsilon $ and 
\begin{equation}\label{eq:Cof}
    C_{\bar{F} \bar{F} } 
=
    \frac{\xi + 1}{2}, \quad 
    C_{ \bar{F}  \bar{f}} = 1, \quad 
    C_{\bar{f} \bar{f}}=
\frac{7 -3 \xi}{6}.
\end{equation}
Using the on-shell condition $ \bar{F}_{\mu \nu}^{a} = f_{\mu \nu}^{a} ( \bar{A} )$,  the expression \eqref{eq:old224} reduces to 
\begin{equation}\label{eq:old226}
    \left . \mathcal{L}^{\text{I}}_{\text{CT}} \right |_{ \bar{F} = f( \bar{A} )} = - \frac{1}{2} d_{N} \frac{11}{3} f_{\mu \nu}^{a} ( \bar{A} ) f^{a \, \mu \nu} ( \bar{A} )
\end{equation}
which agrees with the counterterm Lagrangian obtained in the second order formalism (see Ref.~\cite{abbott:1982}).

One can also verify that the 2-point Green's function with external $ \bar{A} $ fields, are equivalent. In terms of the self-energies shown in Fig.~\ref{fig:BSEFOYM}, such an equivalence of the propagators  may be written in a compact form, as 
\begin{equation}\label{eq:old227}
    D_{\bar{A}\bar{F}} \Pi^{\text{I}}_{\bar{F}\bar{F}} D_{\bar{F}\bar{A}} + 2 D_{\bar{A}\bar{F}} \Pi^{\text{I}}_{\bar{F}\bar{A}} D_{\bar{A}\bar{A}} + D_{\bar{A}\bar{A}} \Pi^{\text{I}}_{\bar{A}\bar{A}} D_{\bar{A}\bar{A}} = D_{\bar{A}\bar{A}} \Pi^{\text{II}}_{\bar{A}\bar{A}} D_{\bar{A}\bar{A}} ,
\end{equation}
where the tree propagators $D_{\bar{x} \bar{y}}$ are the same obtained for the quantum fields, the self-energies in first order formalism are given in Fig.~\ref{fig:BSEFOYM} and $ \Pi^{\text{II}}_{\bar{A} \bar{A}} $ is the gluon self-energy in  second order formalism (see Eqs. 3.3.1--3 of Ref.~\cite{abbott:1982}). 
A straightforward calculation shows that the above propagators are transverse and that \eqref{eq:old227} leads in momentum space to the  equation       
\begin{equation}\label{eq:old228}
    \left ( C_{\bar{F} \bar{F} } + 2 C_{ \bar{F}  \bar{f}} + C_{\bar{f} \bar{f}}\right ) \left ( k_{\mu} k_{\nu} - \eta_{\mu \nu} k^{2}\right ) 
    = \frac{11}{3} ( k_{\mu} k_{\nu} - \eta_{\mu \nu} k^{2} ),
\end{equation}
which is satisfied due to the relations \eqref{eq:Cof}.

\section{Background Ward identities} \label{sec:Ward}

The background effective action $ \bar{\Gamma} ( \bar{\mathfrak{h}}^{\mu \nu}, \tensor{\bar{G}}{_{\mu\nu}^{\lambda}} ) $ defined as the effective action in Eq.~\eqref{eq:213} evaluated for vanishing mean quantum fields is invariant under the background gauge transformations~\eqref{eq:27} \cite{Lavrov:2019nuz}.
This implies that
\begin{equation}\label{eq:ccf55}
    \Delta \bar{\Gamma} ( \bar{\mathfrak{h}}^{\mu \nu}, \tensor{\bar{G}}{_{\mu\nu}^{\lambda}} ) =  \int \mathop{d^4 x} \left(\Delta \bar{\mathfrak{h}}^{\mu \nu} \frac{\delta \bar{\Gamma} }{\delta \bar{\mathfrak{h}}^{\mu \nu} } + \Delta \tensor{\bar{G}}{_{\mu \nu}^{\lambda}} \frac{\delta \bar{\Gamma}}{\delta \tensor{\bar{G}}{_{\mu \nu}^{\lambda}}}  \right)=0.
\end{equation}
This relation leads to Ward identities that reflect the background gauge symmetry of this action.
Such  identities are obtained by taking functional derivatives of Eq.~\eqref{eq:ccf55} with respect to the background fields  $ \bar{\mathfrak{h}}^{\alpha \beta } $ and $ \tensor{\bar{G}}{_{\alpha \beta }^{\lambda}}  $, evaluated at vanishing background fields. 

Replacing Eq.~\eqref{eq:27} in Eq.~\eqref{eq:ccf55}, we have 
\begin{equation}\label{eq:ccf55alt}
    \begin{split}
    & \left[
        \frac{1}{\kappa } (\eta^{\lambda \nu} \partial_{\lambda} \zeta^{\mu} + \eta^{\mu \lambda} \partial_{\lambda} \zeta^{\nu} - \eta^{\mu \nu}\partial_{\lambda}   \zeta^{\lambda} )
   + \bar{\mathfrak{h}}^{\lambda \nu} \partial_{\lambda} \zeta^{\mu} + \bar{\mathfrak{h}}^{\mu \lambda} \partial_{\lambda} \zeta^{\nu} - \partial_{\lambda} ( \bar{\mathfrak{h}}^{\mu \nu} \zeta^{\lambda} )
    \right] 
    \frac{\delta \bar{\Gamma} }{\delta \bar{\mathfrak{h}}^{\mu \nu} } \\ & + \left[
        - \partial_{\mu} \partial_{\nu}  \zeta^{\lambda} + \frac{1}{2} ( \delta_{\mu}^{\lambda} \partial_{\nu} + \delta_{\nu}^{\lambda } \partial_{\mu} ) \partial_{\rho} \zeta^{\rho}  - \zeta^{\rho} \partial_{\rho} \tensor{\bar{G}}{_{\mu \nu}^{\lambda}} + \tensor{\bar{G}}{_{\mu \nu}^{\rho}} \partial_{\rho} \zeta^{\lambda} -  ( \tensor{\bar{G}}{_{\mu \rho}^{\lambda}} \partial_{\nu} + \tensor{\bar{G}}{_{\nu \rho}^{\lambda}} \partial_{\mu}  ) \zeta^{\rho}
    \right]\frac{\delta \bar{\Gamma}}{\delta \tensor{\bar{G}}{_{\mu \nu}^{\lambda}}}  =0.
\end{split}
\end{equation}
For instance, taking a derivative with respect to the background metric $ \bar{\mathfrak{h}}^{\alpha \beta} $, it is straightforward to derive the following relation between the self-energies $ \bar{\mathfrak{h}} \bar{\mathfrak{h}}$ and $ \bar{G} \bar{\mathfrak{h}} $:
\begin{equation}\label{eq:ID1}
    i (k^{\mu} \eta^{\nu \rho} + k^{\nu} \eta^{\mu \rho } - k^{\rho} \eta^{\mu \nu} ){\Pi}_{\mu \nu \alpha \beta} (k) 
- 
\kappa \left[k^{\mu} k^{\nu} \delta_{\lambda}^{\rho} - \frac{k^{\rho} }{2} \left( k^{\mu} \delta_{\lambda}^{\nu} + k^{\nu} \delta_{\lambda }^{\mu} \right)\right]{\Pi}_{\mu \nu}{}^{\lambda}{}_{ \alpha \beta}(k)=0.
\end{equation}
Now, taking the derivative with respect to the field $ \tensor{\bar{G}}{_{\alpha \beta}^{\gamma}}$, one obtains
\begin{equation}\label{eq:ID2}
    i (k^{\mu} \eta^{\nu \rho} + k^{\nu} \eta^{\mu \rho } - k^{\rho} \eta^{\mu \nu} )
    {\Pi}_{\mu \nu \alpha \beta}{}^{\gamma}  (k) 
- 
\kappa  \left[k^{\mu} k^{\nu} \delta_{\lambda}^{\rho} - \frac{k^{\rho} }{2} \left( k^{\mu} \delta_{\lambda}^{\nu} + k^{\nu} \delta_{\lambda }^{\mu} \right)\right]
{\Pi}_{\mu \nu}{}^{\lambda}{}_{ \alpha \beta}{}^{\gamma } (k) =0,
\end{equation}
which relates the self-energies $ \bar{\mathfrak{h}} \bar{G} $ and $ \bar{G} \bar{G} $.
To derive these identities, we have used the fact that single functional derivatives $ \delta \bar{\Gamma} / \delta \bar{\mathfrak{h}^{\mu \nu}} $ and $ \delta \bar{\Gamma} / \delta \tensor{\bar{G}}{_{\mu \nu}^{\lambda}}$ (tadpoles) vanish in our theory. Both identities were explicitly verified at tree and one-loop level.

\subsection{On-Shell Ward identities}

Using the equation of motion \eqref{eq:29}, the Eq.~\eqref{eq:ccf55} reduces to 
\begin{equation}\label{eq:secondccf}
    \Delta \bar{\Gamma} \left( \bar{\mathfrak{h}}^{\mu \nu},  \mathcal{G} ( \bar{\mathfrak{h}} )\right) 
    = \int \mathop{d^4 x} \Delta \bar{\mathfrak{h}}^{\mu \nu} \frac{\delta \Gamma^{\text{I}} }{\delta \bar{\mathfrak{h}}^{\mu \nu} }=0,
\end{equation}
where $  
\bar{\Gamma} \left( \bar{\mathfrak{h}}^{\mu \nu},  \mathcal{G} ( \bar{\mathfrak{h}} )   \right) \equiv \Gamma^{\text{I}}= \int \mathop{d x} 
\mathcal{L}^{\text{I}}_{\text{eff} } \left( \bar{\mathfrak{h}}^{\mu \nu},  \mathcal{G} ( \bar{\mathfrak{h}} ) \right)$ 
is the proper effective action. Taking a derivative of the above equation with respect to $ \bar{\mathfrak{h}}^{\alpha \beta} $ (and setting it to zero) yields the following expression 
\begin{equation}\label{eq:34ainA}
( \eta^{\mu \rho} k^{\nu} + \eta^{\nu \rho} k^{\mu} - \eta^{\mu \nu} k^{\rho} ) 
    \Pi^{\text{I}}_{\mu \nu \alpha \beta} (k) =0.
\end{equation}
This is the Ward identity satisfied by the proper self-energy (see Eq.~\eqref{eq:34a}). 

One also can show that Eq.~\eqref{eq:34ainA} is a direct consequence of the Ward identities~\eqref{eq:ID1} and \eqref{eq:ID2}. First, let us consider the expression~\eqref{eq:ccf55}, when the independent field \( \tensor{\bar{G}}{_{\mu \nu}^{\lambda}} \) is on-shell: 
\begin{equation}\label{eq:ccf55alte}
    \Delta \bar{\Gamma} \left( \bar{\mathfrak{h}}^{\mu \nu}, \bar{G} = \mathcal{G} \right) 
    =  
    \int \mathop{d^{4}x} \Delta \bar{\mathfrak{h}}^{\mu \nu} \left(
    \frac{\delta}{\delta \bar{\mathfrak{h}}^{\mu \nu}} 
    + 
    \frac{\delta \tensor{\mathcal{G}}{_{\alpha \beta}^{\lambda}}}{\delta \bar{\mathfrak{h}}^{\mu \nu}} 
    \frac{\delta}{\delta \tensor{\mathcal{G}}{_{\alpha \beta}^{\lambda}}} 
    \right) 
    \bar{\Gamma} = 0.
\end{equation}

We define the operator
\begin{equation}\label{eq:opW}
    \mathsf{F}_{\mu \nu} \equiv 
    \frac{\delta}{\delta \bar{\mathfrak{h}}^{\mu \nu}} 
    + 
    \frac{\delta \tensor{\mathcal{G}}{_{\alpha \beta}^{\lambda}}}{\delta \bar{\mathfrak{h}}^{\mu \nu}} 
    \frac{\delta}{\delta \tensor{\mathcal{G}}{_{\alpha \beta}^{\lambda}}}
\end{equation}
which encodes the relation between \( \bar{\Gamma} \) and \( \Gamma^{\text{I}} \):
\begin{equation}\label{eq:relation1}
    \mathsf{F}_{\mu \nu} \bar{\Gamma} = 
    \frac{\delta}{\delta \bar{\mathfrak{h}}^{\mu \nu}} 
    \Gamma^{\text{I}}.
\end{equation}

The Ward identities \eqref{eq:ID1} and \eqref{eq:ID2} correspond to the action of the operators
\begin{equation}\label{eq:opid1}
    \frac{\delta}{\delta \bar{\mathfrak{h}}^{\alpha \beta}} \mathsf{F}_{\mu \nu} 
    \quad \text{and} \quad  
    \frac{\delta}{\delta \tensor{\bar{G}}{_{\alpha \beta}^{\gamma}}} \mathsf{F}_{\mu \nu}
\end{equation}
in the action $ \bar{\Gamma} $.
This implies that 
\begin{equation}\label{eq:wardid}
    \int \mathop{d^{4}x} \, \Delta \bar{\mathfrak{h}}^{\mu \nu} 
    \frac{\delta}{\delta \bar{\mathfrak{h}}^{\alpha \beta}} \mathsf{F}_{\mu \nu} \bar{\Gamma} 
    =0 \quad \text{and} \quad 
    \int \mathop{d^{4}x} \, \Delta \bar{\mathfrak{h}}^{\mu \nu} 
    \frac{\delta \tensor{\mathcal{G}}{_{\pi \tau}^{\rho}}}{\delta \bar{\mathfrak{h}}^{\alpha \beta }} 
    \frac{\delta}{\delta \tensor{\bar{G}}{_{\pi \tau}^{\rho}}} \mathsf{F}_{\mu \nu} \bar{\Gamma}  
    =
    0.
\end{equation}
Adding them, we get
\begin{equation}\label{eq:wardfinal}
    \int \mathop{d^{4}x} \, \Delta \bar{\mathfrak{h}}^{\mu \nu} \left(
    \frac{\delta}{\delta \bar{\mathfrak{h}}^{\alpha \beta}} \mathsf{F}_{\mu \nu} 
    + 
    \frac{\delta \tensor{\mathcal{G}}{_{\pi \tau}^{\rho}}}{\delta \bar{\mathfrak{h}}^{\alpha \beta }} 
    \frac{\delta}{\delta \tensor{\mathcal{G}}{_{\pi \tau}^{\rho}}} \mathsf{F}_{\mu \nu} 
    \right)
    =
    \int \mathop{d^{4}x} \, \Delta \bar{\mathfrak{h}}^{\mu \nu} 
    \mathsf{F}_{\alpha \beta} 
    \mathsf{F}_{\mu \nu},
\end{equation}
up to tadpole contributions. Using the relation~\eqref{eq:relation1} and Eq.~\eqref{eq:ccf55alte}, we derive that 
\begin{equation}\label{eq:wardfinal0}
    \int \mathop{d^{4}x} \, \Delta \bar{\mathfrak{h}}^{\mu \nu} 
    \mathsf{F}_{\alpha \beta} 
    \mathsf{F}_{\mu \nu} \bar{\Gamma} 
    =
    \int \mathop{d^{4}x} \, \Delta \bar{\mathfrak{h}}^{\mu \nu} 
    \frac{\delta}{\delta \bar{\mathfrak{h}}^{\alpha \beta}} 
    \frac{\delta}{\delta \bar{\mathfrak{h}}^{\mu \nu}} 
    \Gamma^{\text{I}} =0.
\end{equation}

\subsubsection{‘t Hooft identities} \label{sec:hooft}

Let us consider the generating functional of Green’s functions in the second  order  EH theory 
\begin{equation}\label{eq:h1}
    Z[j]= \int \mathop{\mathcal{D} h^{\mu \nu}} \exp i \int \mathop{d^4 x} \left [ \mathcal{L}_{\text{inv} } + j_{\mu \nu} h^{\mu \nu}\right ],     
\end{equation}
where $ \mathcal{L}_{\text{inv} } $ is the  gauge invariant EH Lagrangian and $ j_{\mu \nu} $ is an external source. 
In order to extract an infinite gauge group factor, we introduce a  functional $ \Delta_{\text{FP}} [h]$ defined as 
\begin{equation}\label{eq:h2}
    \int \mathop{\mathcal{D} \zeta } \Delta_{\text{FP}} [h] \delta 
    \left [ \partial_{\mu} h^{\mu \nu}_{\zeta} - B^{\nu}\right ] 
    = \text{constant}, 
\end{equation}
where $ \Delta_{FP} [h]$ yields the Faddeev-Popov ghost contributions.

Inserting Eq.~\eqref{eq:h2} in Eq.~\eqref{eq:h1}, evaluated at $j_{\mu \nu} =0$, we get  
\begin{equation}\label{eq:h3}
    Z[0] = \int \mathop{\mathcal{D} h^{\mu \nu}} \mathop{\mathcal{D} \zeta} \Delta_{\text{FP}} [h]
    \mathop{\delta} [ \partial_{\mu} h^{\mu \nu}_{\zeta} - B^{\nu} ] 
    \exp i \int \mathop{d^4 x} \mathcal{L}_\text{inv}.
\end{equation}

Next, making an inverse gauge transformation $ h^{\mu \nu} \to h^{\mu \nu}_{\zeta^{-1}}$,   we obtain 
\begin{equation}\label{eq:h4}
     Z[0] = \int \mathop{\mathcal{D} h^{\mu \nu}}  \Delta_{\text{FP}} [h]
     \mathop{\delta}  \left [ \partial_{\mu} h^{\mu \nu} - B^{\nu}\right ] 
    \exp i \int \mathop{d^4 x} \mathcal{L}_\text{inv},
\end{equation}
where we used that $ \Delta_{\text{FP}} [h]$ is $ \zeta $-independent and omitted an irrelevant infinite constant.             
Because $Z[0]$ is a gauge invariant quantity,  it must be  independent of $B^{\nu} $. Hence, integrating  Eq.~\eqref{eq:h4} over  $ B^{\nu} $ with a gauge-breaking gaussian weight function   
\begin{equation}\label{eq:h5}
    \rho = \exp \left [ -\frac{1}{2 \xi \kappa^{2}} \int \mathop{d^4 x} (B^{\nu} - J^{\nu} )^{2}\right ],
\end{equation}
where $ \xi $ is a  gauge parameter and $ J^{\nu} (x)$ is an arbitrary function, we obtain the result 
\begin{equation}\label{eq:h6}
      Z[0] = \int \mathop{\mathcal{D} h^{\mu \nu}}  \Delta_{\text{FP}} [h]
      \exp i \int \mathop{d^4 x} \left[\mathcal{L}_\text{inv} - \frac{1}{2 \xi \kappa^{2} } \left(\partial_{\mu} h^{\mu \nu} -J^{\nu} \right)^{2}\right]
\end{equation}
which is, in fact, independent of $ J^{\nu} $. 

Thus, expanding  Eq.~\eqref{eq:h6} in powers of $ J^{\mu} (x) J^{\nu} (y)$, all coefficients  of $ J^{\mu} $ must vanish. For instance, setting the coefficient of $ J^{\mu} (x) J^{\nu} (y)$ to zero, we obtain the identity 
\begin{equation}\label{eq:h7}
    \frac{2}{\xi \kappa^{2}} \frac{\partial}{\partial x^{\rho}} \frac{\partial}{\partial y^{\sigma}} \langle 0|T h^{\rho \mu} (x)h^{\rho \nu} (y)| 0 \rangle = \eta^{\mu \nu} \delta^{4} (x-y). 
\end{equation}
This relation can be easily verified at tree level by using the free graviton propagator $ \mathcal{D}_{\mu \rho \gamma \beta} $. Eq.~\eqref{eq:h7} may be written at one-loop order, in the form 
\begin{equation}\label{eq:h8}
    k_{\mu} k_{\nu} \mathcal{D}_{\mu \rho \gamma \beta} (k) \Pi^{\gamma \beta \gamma ' \beta '} (k) \mathcal{D}_{\gamma ' \beta ' \nu \rho} (k) =0, 
\end{equation}
where $ \Pi^{\gamma \beta \gamma ' \beta '} (k) $ is the graviton self-energy. This equation implies the relation 
\begin{equation}\label{eq:h9}
    ( \eta_{\mu \rho} k_{\nu} + \eta_{\nu \rho} k_{\mu} - \eta_{\mu \nu} k_{\rho} )
    \Pi^{\mu \nu \alpha \beta}(k) 
    ( \eta_{\alpha \sigma} k_{\beta} + \eta_{\beta \sigma} k_{\alpha} - \eta_{\alpha \beta} k_{\sigma} )=0.
\end{equation}
We also have verified that the proper self-energy computed in Appendix~\ref{sec:OneLoop} satisfies the 't Hooft identity~\eqref{eq:h9}.

\section{Counterterms in the second order formulation} \label{sec:Shapiro}

In Ref.~\cite{Goncalves:2017jxq}, the counterterms of the EH theory in second order form are computed for a general reparametrization of the metric field
\begin{equation}\label{eq:redField}
    g_{\alpha\beta}' = \left[ 
\bar{g}_{\alpha\beta} 
+ \kappa (\gamma_1 \phi_{\alpha\beta} + \gamma_2 \phi\, \bar{g}_{\alpha\beta}) 
+ \kappa^2 \left( 
\gamma_3 \phi_{\alpha\rho} \phi^\rho{}_\beta 
+ \gamma_4 \phi_{\rho\omega} \phi^{\rho\omega} \bar{g}_{\alpha\beta} 
+ \gamma_5 \phi\, \phi_{\alpha\beta} 
+ \gamma_6 \phi^2 \bar{g}_{\alpha\beta} 
\right) 
\right].
\end{equation}
They find that the divergent part of the counterterms is given by (we set $ \Lambda = 0$, and we recall that we work in $ {{D}}= 4 - 2 \epsilon $)
\begin{equation}\label{eq:divref}
    \Gamma^{(1)}_{\text{div}} = -\mu^{{{D}}-4} \frac{1}{16 \pi^{2}  \epsilon} \int d^4x\, \sqrt{- \bar{g} } \left\{
        g_1 \bar{R}_{\mu \nu}^{2}  + \left(\frac{g_4}{2} - \frac{1}{3} g_{1} \right) \bar{R} ^2 + \frac{g_1 + g_{2}}{2} E_4 + \frac{g_3}{2} \Box \bar{R} ,
\right\}
\end{equation}
where $ \mu$ is an arbitrary scale factor, $ \phi = \phi_{\alpha}^{\alpha} $,  $\Box = \bar{\mathsf{D}}_{\mu} \bar{\mathsf{D}}^{\mu} $ and  \begin{equation} \label{eq:gaussbonnet}
E_{4} =  \bar{R}_{\mu\nu\rho\sigma} \bar{R}^{\mu\nu\rho\sigma} - 4 \bar{R}_{\mu\nu} \bar{R}^{\mu\nu} + \bar{R}^2 \end{equation}  is the Gauss-Bonnet term.  In ${{D}}=4$, $E_4$ vanishes identically. The last term is a total derivative, and it also vanishes.

Thus, the counterterm Lagrangian for the general metric field in Eq.~\eqref{eq:redField} reads 
\begin{equation}\label{eq:CT}
    \mathcal{L}_{\text{CT}}^{\text{II}} = \frac{\sqrt{- \bar{g}}}{16 \pi^{2} \epsilon} \left [ g_1 \bar{R}_{\mu \nu}^{2}  + \left(\frac{g_4}{2} - \frac{g_{1}}{3}  \right) \bar{R} ^2\right ], 
\end{equation}
where
\begin{equation}\label{eq:coeff}
    g_1 ={} \frac{7}{20} + \frac{4\gamma_3^2}{\gamma_1^4} - \frac{2A^2}{\gamma_1^2 B^2}, \quad  
    g_4 ={}\frac{1}{4} - \frac{9E}{2} + \frac{31\gamma_3^2 + 216\gamma_4(\gamma_3 + 2\gamma_4)}{6\gamma_1^4} - \frac{A^2}{3\gamma_1^2 B^2} - \frac{C}{6B^2} + \frac{C^2}{2B^4}, \\
\end{equation}
and
\begin{equation}\label{eq:coeff2}
    A ={}\gamma_3 + 2\gamma_5  , \quad 
    B ={}\gamma_1 + 4\gamma_2 , \quad 
    C ={}  (\gamma_3 + 4\gamma_4) + 4(\gamma_5 + 4\gamma_6), \quad   
    E ={} \frac{\gamma_3 + 4\gamma_4}{\gamma_1^2}.
\end{equation}

\subsection{'t Hooft and Veltman metric}
Using the usual metric field definition, that is,
 \begin{equation}\label{eq:coefmetric}
      \gamma_{1} ={}1, \quad \gamma_{2} = 
     \gamma_{3} =\gamma_{4} =\gamma_{5} = \gamma_{6}= 0;
 \end{equation}
 one obtains the well-known result of 't Hooft and Veltman \cite{tHooft:1974toh}:
\begin{equation}\label{eq:new38}
        \frac{\sqrt{- \bar{g}} }{16 \pi^{2} \epsilon} \left [ \frac{1}{120}
         \bar{R}^{2}    + \frac{7}{20}
\bar{R}_{\mu \nu} \bar{R}^{\mu \nu}\right ].
\end{equation}

\subsection{Goldberg metric }
In this work, we use the Goldberg metric field $ h^{\mu \nu} = \sqrt{-g} g^{\mu \nu} $. Under the expansion of the metric as $ g^{\mu \nu} \to \bar{g}^{\mu \nu} + \kappa \phi^{\mu \nu} $, we get  
\begin{equation}\label{eq:goldred}
    h_{\alpha\beta}' =  \frac{1}{\sqrt{- g}} g_{\alpha \beta} = \frac{1}{\sqrt{- \bar{g}}} \left ( 1 + \frac{\kappa}{2} \phi  +\frac{\kappa^{2} }{8} \phi^{2}-  \frac{1}{4} \phi_{\mu \nu} \phi^{\mu \nu}  + \cdots \right ) \left ( \bar{g}_{\alpha \beta} - \kappa \phi_{\alpha \beta} + \kappa^{2} \phi_{\alpha \mu} \phi^{\mu}_{\beta} + \cdots\right ).
 \end{equation}
 Comparing this with Eq.~\eqref{eq:redField}, we identify:
 \begin{equation}\label{eq:coefGold}
     \gamma_{1} ={}-1, \quad \gamma_{2} ={} \frac{1}{2}, \quad 
     \gamma_{3} ={}1, \quad \gamma_{4} ={} -\frac{1}{4}, \quad  
     \gamma_{5} ={}- \frac{1}{2} \quad \text{and} \quad \gamma_{6} ={} \frac{1}{8}.
 \end{equation}

Using these parameters, we obtain the counterterm Lagrangian for the Goldberg metric in the second order formulation
\begin{equation}\label{eq:38insecond}
    \mathcal{L}_{\text{CT}}^{\text{II}} 
    = 
        \frac{\sqrt{- \bar{g}} }{16 \pi^{2} \epsilon} \left [ -  
         \frac{119}{120}
         \bar{R}^{2}    + \frac{87}{20}
\bar{R}_{\mu \nu} \bar{R}^{\mu \nu}\right ].
\end{equation}
This result is valid only for \( \xi = 1 \), corresponding to the De Donder–Feynman gauge. This is a special case of the general result in Eq.~\eqref{eq:38}.

 Then, we see that the counterterms are both gauge and parametrization dependent. However, the on-shell counterterm Lagrangian must be independent of both, as discussed in Ref.~\cite{Goncalves:2017jxq}. Indeed, when Eq.~\eqref{eq:CT} is evaluated on-shell, we find that
\begin{equation}\label{eq:CTOS}
    \mathcal{L}^{\text{II}}_{\text{CT}} |_{\text{on-shell}} = 0.
\end{equation}

\bibliography{backgroundEH.bib}      
\end{document}